\documentclass[12pt]{article}
\usepackage{epsf}
\usepackage{amsmath}
\usepackage{amsfonts}
\usepackage{psfrag}

\def \IC{\mathbb{C}}
\def \IZ{\mathbb{Z}}
\def \IP{\mathbb{P}}

\def \IF{\mathbb{F}}

\begin{document}
\setlength{\parskip}{2ex} \setlength{\parindent}{0em}
\setlength{\baselineskip}{3ex}
\newcommand{\onefigure}[2]{\begin{figure}[htbp]
         \caption{\small #2\label{#1}(#1)}
         \end{figure}}
\newcommand{\onefigurenocap}[1]{\begin{figure}[h]
         \begin{center}\leavevmode\epsfbox{#1.eps}\end{center}
         \end{figure}}
\renewcommand{\onefigure}[2]{\begin{figure}[htbp]
         \begin{center}\leavevmode\epsfbox{#1.eps}\end{center}
         \caption{\small #2\label{#1}}
         \end{figure}}
\newcommand{\comment}[1]{}
\newcommand{\myref}[1]{(\ref{#1})}
\newcommand{\secref}[1]{sec.~\protect\ref{#1}}
\newcommand{\figref}[1]{Fig.~\protect\ref{#1}}
\newcommand{\mathbold}[1]{\mbox{\boldmath $\bf#1$}}
\newcommand{\mJ}{\mathbold{J}}
\newcommand{\momega}{\mathbold{\omega}}
\newcommand{\bz}{{\bf z}}
\def\bbbz{{\sf Z\!\!\!Z}}
\newcommand{\PP}{\mbox{I}\!\mbox{P}}
\newcommand{\ff}{\mbox{I}\!\mbox{F}}
\newcommand{\bbbc}{\mbox{C}\!\!\!\mbox{I}}
\def\sl2z{SL(2,\IZ)}
\newcommand{\bbbq}{I\!\!Q}
\newcommand{\be}{\begin{equation}}
\newcommand{\ee}{\end{equation}}
\newcommand{\bea}{\begin{eqnarray}}
\newcommand{\eea}{\end{eqnarray}}
\newcommand{\nn}{\nonumber}
\newcommand{\unit}{1\!\!1}
\newcommand{\half}{\frac{1}{2}}
\newcommand{\shalf}{\mbox{$\half$}}
\newcommand{\transform}[1]{
   \stackrel{#1}{-\hspace{-1.2ex}-\hspace{-1.2ex}\longrightarrow}}
\newcommand{\inter}[2]{\null^{\#}(#1\cdot#2)}
\newcommand{\lprod}[2]{\vec{#1}\cdot\vec{#2}}
\newcommand{\mult}[1]{{\cal N}(#1)}
\newcommand{\Bn}{{\cal B}_N}
\newcommand{\B}{{\cal B}}
\newcommand{\Beight}{{\cal B}_8}
\newcommand{\Bnine}{{\cal B}_9}
\newcommand{\Eman}{\widehat{\cal E}_N}
\newcommand{\C}{{\cal C}}
\newcommand{\Q}{Q\!\!\!Q}
\newcommand{\comp}{C\!\!\!C}
\newdimen\tableauside\tableauside=1.0ex
\newdimen\tableaurule\tableaurule=0.4pt
\newdimen\tableaustep
\def\phantomhrule#1{\hbox{\vbox to0pt{\hrule height\tableaurule width#1\vss}}}
\def\phantomvrule#1{\vbox{\hbox to0pt{\vrule width\tableaurule height#1\hss}}}
\def\sqr{\vbox{%
  \phantomhrule\tableaustep
  \hbox{\phantomvrule\tableaustep\kern\tableaustep\phantomvrule\tableaustep}%
  \hbox{\vbox{\phantomhrule\tableauside}\kern-\tableaurule}}}
\def\squares#1{\hbox{\count0=#1\noindent\loop\sqr
  \advance\count0 by-1 \ifnum\count0>0\repeat}}
\def\tableau#1{\vcenter{\offinterlineskip
  \tableaustep=\tableauside\advance\tableaustep by-\tableaurule
  \kern\normallineskip\hbox
    {\kern\normallineskip\vbox
      {\gettableau#1 0 }%
     \kern\normallineskip\kern\tableaurule}%
  \kern\normallineskip\kern\tableaurule}}
\def\gettableau#1 {\ifnum#1=0\let\next=\null\else
  \squares{#1}\let\next=\gettableau\fi\next}

\tableauside=1.0ex
\tableaurule=0.4pt

\def\IE{\relax{\rm I\kern-.18em E}}
\def\IP{\relax{\rm I\kern-.18em P}}

\noindent
\begin{titlepage}

\begin{center}
\today \hfill HUTP-03/A038\\\hfill SLAC-PUB-9924 \\
\hfill SU-ITP 03/12\\  \hfill hep-th/0306032\\ \vskip 1cm
{\large {\bf  $SU(N)$ Geometries and Topological String Amplitudes}} \vskip 2cm
{Amer Iqbal$^{1}$, Amir-Kian Kashani-Poor$^{2}$}\\ \vskip 0.5cm

{$^{1}$Jefferson Laboratory\\
Harvard University,\\
Cambridge, MA  02138, U.S.A.\\}
\vskip 0.5cm

{$^{2}$ Department of Physics and SLAC\\
Stanford University,\\
Stanford, CA 94305/94309, U.S.A.\\}

\end{center}

\begin{abstract}
It has been conjectured recently that the field theory limit of the topological string partition functions, including all higher genus contributions, for the family of CY3-folds giving rise to ${\cal N}=2$ 4D $SU(N)$ gauge theory 
via geometric engineering can be obtained from gauge instanton calculus. We verify this surprising conjecture by calculating the partition functions for such local CYs using diagrammatic techniques inspired by geometric transitions. Determining the Gopakumar-Vafa invariants for these geometries to all orders in the fiber wrappings allows us to take the field theory limit.
\end{abstract}
\end{titlepage}
\newpage

\thispagestyle{empty}

\pagenumbering{arabic}
\section{Introduction}
Topological string theory has received much attention recently due to its implications for large N dualities
in the physical string theory \cite{vafa1}. The amplitudes of the topological string theory not only have mathematical significance as generating functions of Gromov-Witten invariants but also compute coefficients of certain F-terms in the physical
$4d$ theory. Much progress has been made toward their calculation for local toric CY3-folds in the past year \cite{AMV,DFG,paper,AKMV}.
   
In this paper, we will compute the topological string partition function  $F=\sum {F_g g_s^{2g-2}}$ for  local Calabi-Yau 3-folds which are resolved $A_n$ singularities fibered over $\IP^1$. These geometries are used to geometrically engineer ${\cal N}=2$ $D=4$ $SU(N)$ theories \cite{Klemm:1996bj,KKV,KLMVW,Katz:1997eq}. Our interest in these theories stems from the following recent developments in instanton calculus \cite{Nekrasov,algorithm,Bruzzo:2002xf}. The problem of computing the gauge instanton coefficients ${\cal F}_k$ was reduced to the solution of certain integrals over the (reduced) instanton moduli space in \cite{DKM,Dorey:1996bf,Bellisai:2000tn,Bellisai:2000bc,Flume:2001kb,Hollowood:2002ds}. These integrals prove difficult to solve for $k\ge2$. \cite{Nekrasov} performs a deformation of the integrand which allows their evaluation. The deformed integrals are then assembled into a generating function, the expression for which is computed in \cite{Nekrasov, algorithm,Bruzzo:2002xf}, and the ${\cal F}_k$ can be extracted from this expression. The surprising conjecture in \cite{Nekrasov} is that this generating function itself has an interpretation: it represents the field theory limit of the topological string partition function on the local CY which geometrically engineers the gauge theory! The deformation parameter here plays the role of the string coupling. Inspired by the form of this expression, \cite{Nekrasov} also contains a conjecture about the form of the full string partition function, before taking the field theory limit. It is these two conjectures, the interpretation of the generating function and the form of the full string partition function, which we wish to investigate, and which we verify, the latter for a certain choice of fibration of the geometry, in this paper.

Specifically, the form of the topological string partition function conjectured in \cite{Nekrasov} is
\bea
Z_{Nekrasov}:=\sum_{R_{1,\cdots,N}}\varphi^{l_{R_{1}}+\cdots l_{R_{N}}}
\prod_{l,n=1}^{N}\prod_{i,j=1}^{\infty}\frac{\mbox{sinh}\frac{\beta}{2}(a_{ln}+\hbar(\mu_{l,i}-\mu_{n,i}+j-i))}{\mbox{sinh}\frac{\beta}{2}(a_{ln}+\hbar(j-i))}\,.
\eea
This turns out to be the topological string partition function of the distinguished fibration of resolved $A_{N-1}$ over $\IP^{1}$ which is an orbifold of the resolved conifold. The expression we obtain in this 
paper for this case is 
\bea
{\widehat Z} &=& \sum_{R_{1,\cdots,N}} \varphi^{l_{1}+\cdots l_{N}}
\frac{\prod_{i=1}^{N}{\cal W}_{R_{i}}(q)^{2}}{\prod_{1\leq i<j\leq N}\prod_{k}(1-q^{k}Q_{i}\cdots Q_{j-1})^{2C_{k}(R_{i},R_{j}^{T})}} \,,
\eea
where $\varphi$ is a combination of K\"ahler parameters. We demonstrate the equivalence of these two expressions in section \ref{conjecture} of this paper.

The technique we use to calculate the string partition function is inspired by \cite{AMV}. There, it is shown how to pass to an open string geometry dual to the closed geometry, on which the partition function can be calculated using Chern-Simons theory. In \cite{paper}, diagrammatic rules are extracted from this procedure for a subclass of geometries which allow the calculation to proceed without knowledge of the open string geometry. These rules have been completed and given a physical interpretation in the recent paper \cite{AKMV}. 

To perform the calculation of the string partition function in the case at hand, a further hurdle must be overcome. The expression obtained by \cite{Nekrasov, algorithm,Bruzzo:2002xf} and the one conjectured by \cite{Nekrasov} translate into topological string theory expressions which are an expansion in wrappings of the base $\IP^1$ but {\it exact} in fiber wrappings. We show how to obtain these exact results by making an assumption about the form of an expression which enters in the Chern-Simons calculation, generalizing an approach utilized in \cite{firstpaper}.

The plan of the paper is as follows. In section \ref{geometric}, we briefly review how ${\cal N}=2$ $SU(N)$ gauge theories in $D=4$ are engineered in string theory. Section \ref{gtransitions} explains the diagrammatic techniques we employ to perform our calculation, and their origins. In section \ref{allorders}, we introduce the final ingredient of our calculation which allows the determining of the integral invariants to all orders in fiber wrappings, perform the calculation for $SU(3)$ geometries in detail, and show how the result generalizes to $SU(N)$. In the final section, we compare our results to those obtained based on gauge instanton calculus. In the appendix, we provide some details on the geometries studied in this paper and their toric description, and make some comments relating to the $5d$ theory one obtains by considering M-theory on these CY geometries.

\section{Geometric engineering of $SU(N)$ theories} \label{geometric}
Compactifications of type IIA on singular CYs yield effective four dimensional theories with enhanced gauge symmetry \cite{KLMVW,Klemm:1996bj,KKV}. The gauge symmetry 
in the field theory arises from D2-branes wrapping
collapsing curves in the CY3-fold. Thus to get a 
particular gauge symmetry one has to study a 
CY3-fold with the appropriate shrinking cycles.

\begin{figure}[h]
\begin{center}\leavevmode\epsfbox{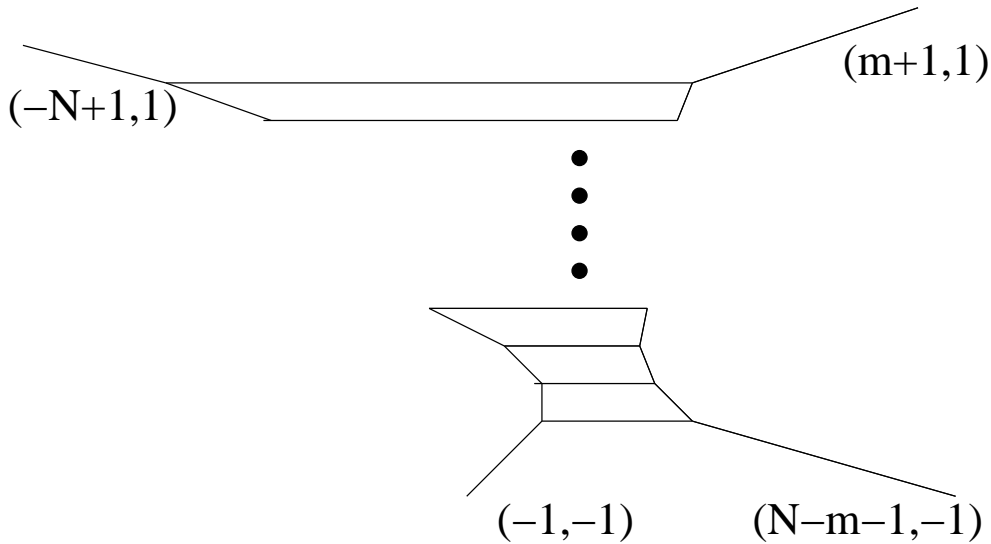}\end{center} 
\caption{\small The web diagram for $SU(N)$ geometries. The tupels in parentheses signify the slope of the respective lines. \label{sunweb}}
\end{figure}

The engineering of an $SU(N)$ gauge theory requires
a singularity of $A_{N-1}$ type. Type IIA compactification on such a geometry
gives a six dimensional $SU(N)$ theory with sixteen supercharges. To obtain
a four dimensional theory further compactification on a two dimensional surface
is required. If the four dimensional surface is $T^{2}$ the four dimensional theory acquires ${\cal N}=4$ supersymmetry. To break supersymmetry down to
${\cal N}=2$ (eight supercharges) the surface should have no holomorphic
one forms and therefore has to be a $\IP^{1}$. To obtain a CY3-fold, the $A_{N-1}$ must be fibered non-trivially over the $\IP^{1}$. The web diagram corresponding to such a geometry is given in \figref{sunweb}.
We review aspects of such geometries in the appendix. The details of the ${\cal N}=2$ theory obtained by type IIA compactification on such a CY3-fold depend on the choice of fibration. In the field theory limit, which we review next, all such 3-folds reduce to the same theory.

The field theory limit is obtained by taking the string scale to infinity. By the relations of the base and fiber K\"ahler parameters to the gauge coupling and W-boson masses, these parameters must be scaled as\footnote{In the following, the notation $Q_c$ will always be reserved for the exponential of minus the corresponding K\"ahler parameter, $e^{-T_c}$.}
\begin{eqnarray} \label{fieldtheorylimit}
Q_{B}:=e^{-T_{B}}=(\frac{\beta \Lambda}{2})^{2N}\,,\,\,\,Q_{F_{i}}:=e^{-T_{F_{i}}}=e^{-\beta a_{i,i+1}}\,\,i=1,\cdots,N-1\,. 
\end{eqnarray}

$\Lambda$ in the above denotes the quantum scale in four 
dimensions, the $a_{i,i+1}=a_{i+1}-a_i$ parameterize the VEVs of adjoint scalars in the Cartan subalgebra of the gauge group, and the parameter $\beta$ is introduced such that the field theory limit corresponds to $\beta\rightarrow 0$. 

The ${\cal N}=2$ prepotential has both 1-loop perturbative and non-perturbative (instanton) contributions,
\bea
{\cal F}={\cal F}_{classical}+{\cal F}_{1-loop}+\sum_{k=1}^{\infty}
c_{k}(a_{i})\Lambda^{2Nk}\,.
\eea
We compare this to the expansion of the genus zero topological 
string amplitude
\bea
F_{0}(T_{B},\{T_{F_{i}}\})&=&P_{3}(T_{B},\{T_{F_{i}}\})+\sum_{(k,{\bold m})\neq (0,{\bold 0})}
\sum_{n=1}^{\infty}\frac{N^{0}_{(k,{\bold m})}}{n^{3}}
e^{-nkT_{B}-n\sum_{i}m_{i}T_{F_{i}}}\,,
\eea
(here $P_{3}(T_{B},\{T_{F_{i}}\})$ is a cubic polynomial from which one gets the classical
contribution to the prepotential). The contributions of worldsheet instanton multiwrappings, $n>1$, vanish in the field theory limit. By considering (\ref{fieldtheorylimit}), it then becomes clear that the $k$-th gauge instanton contributions stem from worldsheet instantons that wrap the base $\IP^1$ of our geometries $k$-times.

In this paper, we will be interested in taking the field theory limit of the full topological partition function $\sum g_s^{2g-2} F_g$, rather than just studying the genus 0 contribution. We are motivated to study the full quantity due to recent works  \cite{Nekrasov,algorithm,Bruzzo:2002xf} which obtain it, as reviewed in section \ref{conjecture}, via instanton calculations within field theory. Obtaining finite contributions from all genera requires scaling the string coupling such that $q:=e^{i g_s}=e^{\beta \hbar}$. $\hbar$ will serve to distinguish between the contributions at different $g_s$ (the notation is chosen in accordance with \cite{Nekrasov}).

\section{Diagrammatics} \label{gtransitions}
\subsection{Geometric transitions}
In \cite{GV}, the string theory partition function $\sum g_s^{2g-2} F_g$ is shown to have the following form
\begin{eqnarray} \label{gvpfsin}
\sum_{g=0}^{\infty}g_{s}^{2g-2}F_{g}(\omega)=\sum_{\Sigma\in H_{2}(X)}\sum_{g=0}^{\infty}\sum_{n=1}^{\infty}
\frac{{N}^{g}_{\Sigma}}{n} (2\sin(n\frac{ g_s}{2}))^{2g-2}\,e^{-n\Sigma\cdot \omega}\,.
\end{eqnarray}
In \cite{AMV}, the Gopakumar-Vafa invariants ${N}^{g}_{\Sigma}$ for a given local Calabi-Yau are determined by using duality to the open topological string on a deformed geometry obtained by performing local conifold transitions. The worldsheet instantons of this deformed geometry are under strict control, and by \cite{WCS}, Chern-Simons theory can be used to determine the open string partition function. In particular, open string worldsheet instantons map the boundaries of the worldsheet to $S^3$'s in the target space, and their contribution to the partition function is captured within the Chern-Simons theory by Wilson loops on the image of these boundaries. For this paper, we will require the expression $W_{R_1 R_2}$ for the expectation value of two Wilson loops in representations $R_1$, $R_2$ of $SU(N)$ on an $S^3$ forming a Hopf link. This is given by\footnote{For a derivation of the following expressions for Wilson loop amplitudes in Chern-Simons theory the reader is referred to \cite{marino} and references cited therein.}
\begin{eqnarray} \label{wr1r2}
W_{R_1 R_2}&=&\mbox{dim}_q R_1 (\lambda q)^{\frac{l_2}{2}} s_{\mu^2}(E_{\mu^1}(t)) \,.
\end{eqnarray}
Here, $q$ is the exponential of the Chern-Simons coupling, $q=\exp(\frac{2 \pi i }{k+N})$, $\lambda$ the exponential of the 't Hooft coupling, $\lambda=q^N$. $\mu^{1,2}$ denote the Young tableaux corresponding to the representations $R_1$, $R_2$. $\mbox{dim}_q R$, the quantum dimension of the representation $R$, is the normalized expectation value of a Wilson loop in representation $R$ on an unknot, given by 
\begin{eqnarray}
\mbox{dim}_q R &=&\prod_{1\leq i<j\leq d}\frac{[\mu_{i}-\mu_{j}+j-i]}{[j-i]}\,\prod_{i=1}^{d}\prod_{v=1}^{\mu_{i}}\frac{[v]_\lambda}{[v-i+d]} \,,
\end{eqnarray}
where $[x]_\lambda=\lambda^{\frac{1}{2}}q^{x/2}-\lambda^{-\frac{1}{2}} q^{-x/2}$, $[x] = [x]_1$, $d$ denotes the number of rows in the tableau $\mu$ and $\mu_i$ denotes the number of boxes in the $i$-th row of $\mu$. Finally, $s_\mu$ is the Schur polynomial of the representation 
described by $\mu$, given by
\begin{eqnarray}
s_\mu &=& \det M_\mu \,,
\end{eqnarray}
where the $r \times r$ matrix $M_\mu$, with $r$ the number of columns in $\mu$, is given by $M_\mu^{(ij)}=(a_{\mu^{\vee}_{i}+j-i})$. $\mu^{\vee}$ is the transposed Young tableaux to $\mu$, obtained by interchanging columns and rows. The $a_i$ are the coefficients of the power series which is the argument of the Schur polynomial, in our case the coefficients of $t^i$ in the expansion of $E_{\mu}$, given by
\bea
E_{\mu}(t)&=&(1+\sum_{n=1}^{\infty}(\prod_{i=1}^{n}\frac{1-\lambda^{-1} q^{i-1}}{q^{i}-1})t^{n})\,(\prod_{j=1}^{d}\frac{1+q^{\mu_{j}-j}t}{1+q^{-j}t})\,.
\eea
The open string parameters $q$ and $\lambda$ map to the closed string parameters $e^{i g_s}$ and $e^t$, where $t$ is the K\"ahler parameter of the compact curve obtained by resolving the conifold singularity (note that $q$ and $\lambda$ are not independent parameters, whereas $g_s$ and $t$ are). Upon rewriting the Chern-Simons amplitude in terms of closed string parameters, one obtains all Gopakumar-Vafa invariants ${N}^{g}_{\Sigma}$ up to a given degree in $\Sigma$. 

\begin{figure}[p]
\begin{center}\leavevmode\epsfbox{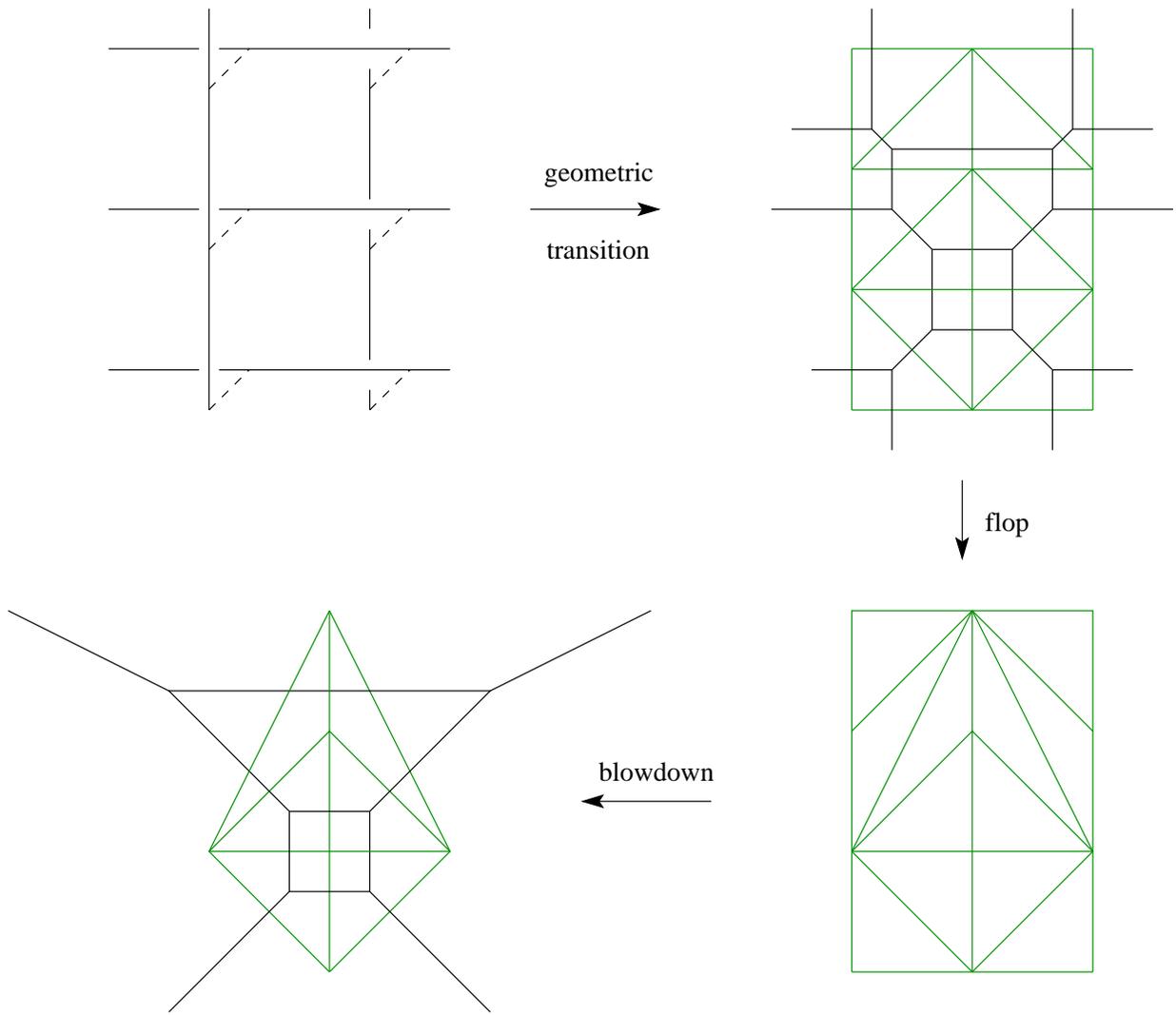}\end{center} 
\caption{\small The open geometry dual to a resolved $A_2$ over $\IP^1$ fibration. \label{transition}}
\end{figure}

The open geometry related via flops and blowdowns to the $SU(3)$ geometry is depicted in \figref{transition}.\footnote{For a more detailed description of such transitions and an explanation of diagrams such as \figref{transition}, see e.g. \cite{firstpaper}.}
As we describe in the next subsection, it is not necessary to compute the complete open string partition function and then take appropriate limits to arrive at the desired closed string result. A shortcut is available.

\subsection{The emergence of diagrammatic rules}
Upon calculating closed topological amplitudes using geometric transitions 
and CS theory, it was noticed by several authors \cite{AMV,paper,AKMV} that 
diagrammatic rules emerge which allow writing down the amplitude by 
considering the web diagram of the closed string geometry. All internal lines are labelled by representations of $SU(N)$, which must be summed over. They contribute factors $Q_c^{l}= e^{-l T_c}$ to the amplitude, where $T_c$ is the K\"ahler class of the curve represented by the internal line, and $l$ is the length of the representation. Vertices with two internal lines carry a factor ${\cal W}_{R_1,R_2}$, which is the leading order contribution in $\lambda$ to the quantity $W_{R_1 R_2}$ introduce in (\ref{wr1r2}) above,
\bea
{\cal W}_{R_{1}R_{2}}(q)={\cal W}_{R_{1}}(q)\,q^{\frac{l_{R_{2}}}{2}}\,
s_{\mu_{R_{2}}}({\cal E}_{\mu_{R_{1}}}(t))\,,
\eea
with
\begin{eqnarray}
{\cal E}_\mu&=& (1+\sum_{n=1}^{\infty}(\prod_{i=1}^{n}\frac{1}{q^{i}-1})t^{n})\,(\prod_{j=1}^{d}\frac{1+q^{\mu_{j}-j}t}{1+q^{-j}t})\,.
\end{eqnarray}
In particular, ${\cal W}_{R} = {\cal W}_{R \cdot}$ is given by 
\bea
{\cal W}_{R}(q)&=&q^{\kappa_{R}/4}\prod_{1\leq i<j\leq d}\frac{[\mu_{i}-\mu_{j}+j-i]}{[j-i]}\,\prod_{i=1}^{d}\prod_{v=1}^{\mu_{i}}\frac{1}{[v-i+d]}\,,
\label{weq}
\eea
where $\kappa_R$ is 
\begin{equation}
\kappa_R=l_R + \sum_{i=1}^{d(\mu)}\mu_i (\mu_i - 2i) \,.
\end{equation}

These rules are inspired by considering the open string dual to an extended closed geometry, related to the original geometry by blow-ups, such that all internal lines of the original geometry correspond to annulus instantons in the open geometry.\footnote{Note that this procedure might lead to additional compact divisors, which manifest themselves in the web diagram as crossing external lines.} The geometric transition gives rise to additional compact curves, one per local transition from a deformed to a resolved conifold, which are eliminated by taking their K\"ahler parameters $\lambda_i$ to infinity. As long as only two annuli end on an $S^3$ in the open geometry, each $S^3$ contributes factors $W_{R_1 R_2}$ to the amplitude, and the $\lambda_i \rightarrow \infty$ limit yields the vertex factors ${\cal W}_{R_1 R_2}$ as claimed above. 

\cite{paper} points out the similarity of the diagrammatic rules to Feynman rules, where the K\"ahler parameters of the 3-fold play the role of Schwinger parameters, the vertices are given by ${\cal W}_{R_i R_j}$ and a framing factor described below, and the factor $e^{-l_{1}r}\delta_{R_{1}R_{2}}$ can be interpreted as a propagator. This approach has recently been made rigorous in \cite{AKMV}.

Let us clarify the diagrammatic approach by looking at two examples.

\begin{figure}[h]
\begin{center}\leavevmode\epsfbox{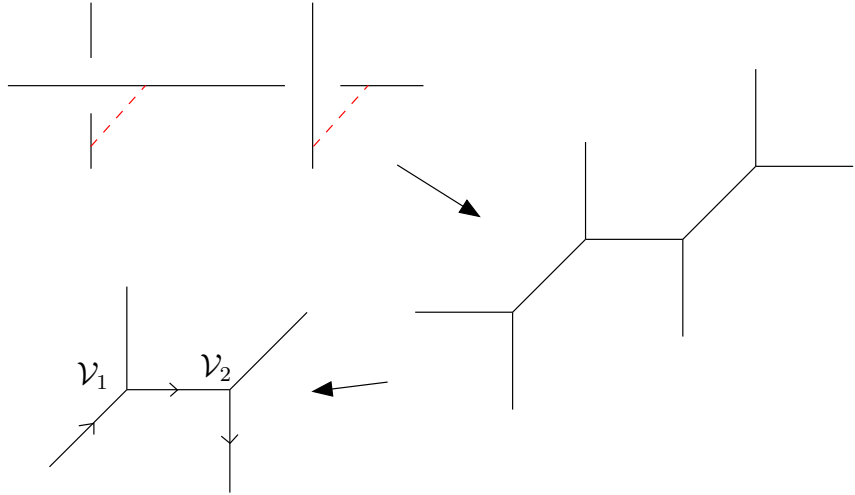}\end{center} 
\caption{\small The path to the resolved conifold partition function which yields diagrammatic rules. \label{diagconf}}
\end{figure}

Consider first the resolved conifold. The relevant amplitude on the open string side here is of course, via the conifold transition, the CS partition function on a sphere. To obtain an expression which adheres to our diagrammatic rules, we follow the seemingly more cumbersome path depicted in \figref{diagconf} to arrive at the partition function of the resolved conifold. This is in accordance with the procedure outlined above of relating compact curves of the closed geometry to annulus instantons in the open geometry.
\psfrag{v1}{${\cal V}_1$}
\psfrag{v2}{${\cal V}_2$}
Equating the two expressions yields the relation
\begin{eqnarray} \label{cfid}
\frac{Z}{S^{-1}_{00}(\lambda_1)S^{-1}_{00}(\lambda_2)}&=& \lim_{\lambda_1,\lambda_2 \rightarrow \infty} \sum W_{\cdot R}(\lambda_1,q)W_{\cdot R}(\lambda_2,q) (\frac{Q}{\sqrt{\lambda_1\lambda_2}})^{l_R} (-1)^{l_{R}}q^{-\frac{\kappa_{R}}{2}}   \\ 
&=&  \sum {\cal W}_{\cdot R}(q){\cal W}_{\cdot R}(q) Q^{l_R} (-1)^{l_{R}}q^{-\frac{\kappa_{R}}{2}} \nn \\
&=& S^{-1}_{00}(Q) \,. \nn
\end{eqnarray}
Note that $Q$ must be renormalized by a factor $\frac{1}{\sqrt{\lambda_1\lambda_2}}$ in order for the $\lambda_1,\lambda_2 \rightarrow \infty$ limit to exist. Note also that the limit yields the CS partition function on $S^3$ proper, not divided by $S^{-1}_{00}$. Finally, the factor $(-1)^{l_{R}}q^{-\frac{\kappa_{R}}{2}}$ is a framing factor. \cite{paper} proposes the following diagrammatic rules to determine these. Associate with each vertex of the web diagram an $\sl2z$ matrix which maps 
the $(p,q)$ charge (the slope) of one leg to the $(p',q')$ charge of the other, with $(p,q)$ and $(p',q')$ being the charges associated with two internal legs. In our example, we only have one internal leg at each vertex, so we must think of the bottom diagram in \figref{diagconf} as embedded in a larger geometry. The matrix $T^{m}S^{-1}T^{n}$ at a vertex ${\cal W}_{R_1 R_2}$ gives rise to a framing factor $(-1)^{n l_1 + m l_2} q^{n\kappa_{1}/2+m\kappa_{2}/2}$.
For the diagram shown, we obtain the following $\sl2z$ matrices,
\begin{eqnarray}
{\cal V}_{1}= S^{-1}T^{-1}\,,\,\,\,{\cal V}_{2}=S^{-1}\,,
\end{eqnarray}
in accordance with the framing factor exhibited in (\ref{cfid}). The expression for $S^{-1}_{00}(Q)$ obtained here will be useful shortly.

\begin{figure}[h]
\begin{center}
\leavevmode\epsfbox{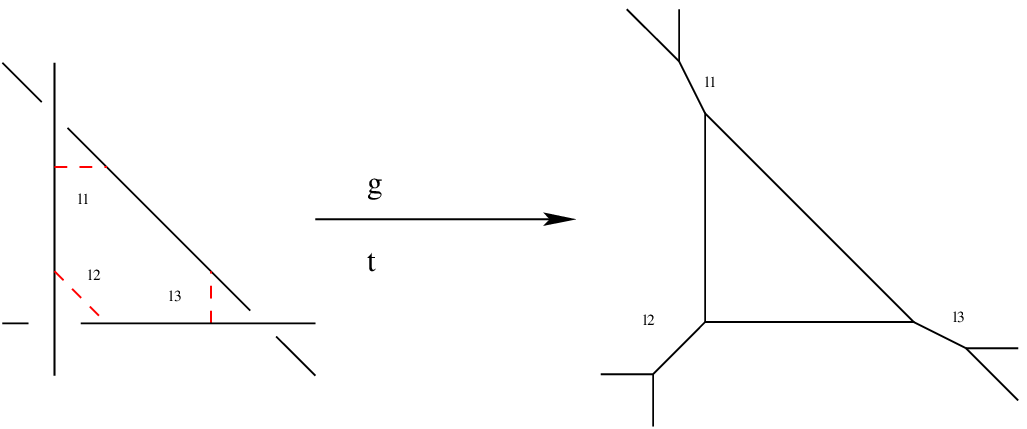}
\end{center} 
\caption{\small Obtaining ${\cal B}_1$ via geometric transition and limits.}\label{transb1}
\end{figure}

\begin{figure}[h]
\begin{center}
\leavevmode\epsfbox{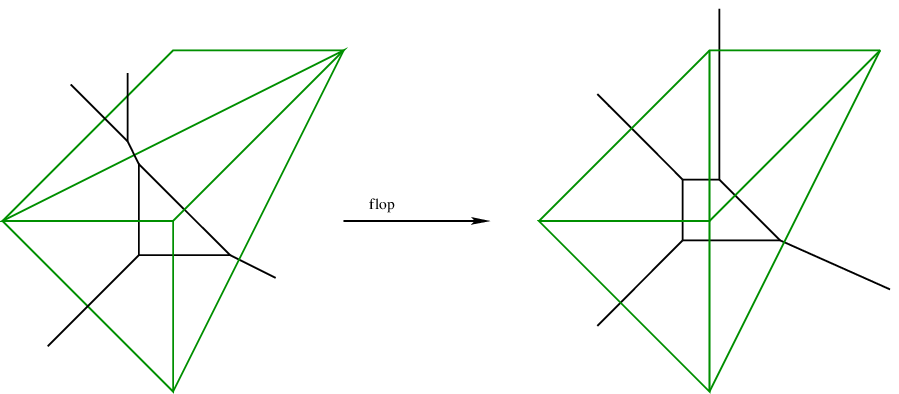}
\end{center} 
\caption{\small The flop relating the limit of the geometry in \figref{transb1} to ${\cal B}_1$.}\label{flop}
\end{figure}

The second example we wish to consider is local $\IP^{2}$ 
blown up at one point, which is the first del Pezzo surface ${\cal B}_1$. The web diagram which can be obtained from the 
toric data is shown in \figref{B1} below. 
\psfrag{r1}{$R_1$}
\psfrag{r2}{$R_2$}
\psfrag{r3}{$R_3$}
\psfrag{r4}{$R_4$}
\psfrag{v3}{${\cal V}_3$}
\psfrag{v4}{${\cal V}_4$}
\onefigure{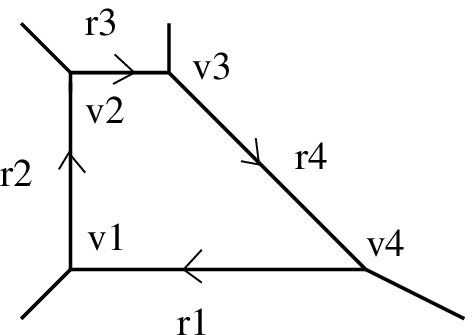}{Web diagram of local $\IP^2$ blown up at one point.}
Again, there are two ways of obtaining the partition function. In \cite{AMV}, see also \cite{DFG}, the geometric transition depicted in \figref{transb1} is considered.
\psfrag{l1}{$\lambda_1$}
\psfrag{l2}{$\lambda_2$}
\psfrag{l3}{$\lambda_3$}
\psfrag{l4}{$\lambda_4$}
\psfrag{g}{geometric}
\psfrag{t}{transition}
Taking the $\lambda_{2,3} \rightarrow \infty$ limit yields a geometry which is related to ${\cal B}_1$ by a flop, see \figref{flop}.
\psfrag{flop}{flop}
In this approach, the K\"ahler class of the exceptional divisor of the del Pezzo is given by $\log \lambda_1$, i.e. related to the exponential of the Chern-Simons coupling in the open picture. Instead, we can apply the diagrammatic rules outlined above to this example. Here, the K\"ahler class of the exceptional divisor is related to the renormalized area of the annulus instanton stretching between vertex ${\cal V}_2$ and ${\cal V}_3$ in \figref{B1}.

The $\sl2z$ matrices associated to the vertices for this example are given by
\bea
{\cal V}_{1}= S^{-1}\,,\,\,\,{\cal V}_{2}=S^{-1}\,,\,\,\,{\cal V}_{3}=T^{-1}S^{-1}T^{-1}\,,\,\,\,{\cal V}_{4}=TS^{-1}T\,.
\eea

We thus obtain for local ${\cal B}_{1}$ 
\bea
Z&=&\sum_{R_{1,2,3,4}}Q_{B}^{l_{1}+l_{3}}Q_{F}^{l_{1}+l_{2}+l_{4}}\,(-1)^{l_{1}+l_{3}}\,q^{\frac{1}{2}(\kappa_{1}-\kappa_{3})}\,{\cal W}_{R_{1},R_{2}}{\cal W}_{R_{2},R_{3}}{\cal W}_{R_{3},R_{4}}{\cal W}_{R_{4},R_{1}}\,,
\label{b1}
\eea
where $T_{B,F}$ are the K\"ahler parameters of the base and the fiber $\IP^{1}$ with
$Q_{B,F}=e^{-T_{B,F}}$.  Note that the expression we obtain from applying diagrammatics only contains ${\cal W}_{R_1 R_2}$'s, which are algebraically simpler than the $W_{R_1 R_2}$ that arises in the first approach to this example described above.

Let us define the following generating function of the
Gopakumar-Vafa invariants,
\bea
f^{(n)}_{g}(x)=\sum_{m}(-1)^{g-1}N^{g}_{(n,m)}x^{m}\,,
\eea
then from (\ref{b1}) it follows that
\bea\nn
f^{(1)}_{g}(x)&=&\delta_{g,0}(1+3x+5x^2+7x^3+9x^4+11x^5+13x^6+15x^7+17x^8+19x^9+21x^{10}+\cdots)\,,\\ \nn
f^{(2)}_{0}(x)&=&6x^2+32x^3+110x^4+288x^5+644x^6+1280x^7+2340x^8+4000x^9+
6490x^{10}+\cdots\,,\\ \nn
f^{(2)}_{1}(x)&=&9x^3+68x^4+300x^5+988x^6+2698x^7+6444x^8+13916x^9+27764x^{10}+\cdots\,,\\\nn
f^{(2)}_{2}(x)&=&12x^4+116x^5+628x^6+2488x^7+836x^8+22404x^9+55836x^{10}+\cdots\,
\eea

\subsection{The three point vertex: ${\cal V}_{R_{1}R_{2}R_{3}}(q)$}
The examples studied above involve open configurations in which at most two annulus instantons end on the same $S^3$, or, in terms of diagrammatics, only vertices with at most two internal lines attached occur. 

As shown in a recent paper \cite{AKMV} an effective vertex on which three internal lines end (we will refer to this as a three point vertex) can be formulated as well. Once the existence of this vertex is established, the expressions for the vertices $V_{.,R_{1},R_{2}}$ and $V_{\tableau{1},R_{1},R_{2}}$ we will need for our computations can easily be determined as follows.\footnote{Our vertex differs slightly from \cite{AKMV} in the choice of framing factors.}

We consider the subdiagram of a web diagram and its transition, as depicted in \figref{vertex}.
\onefigure{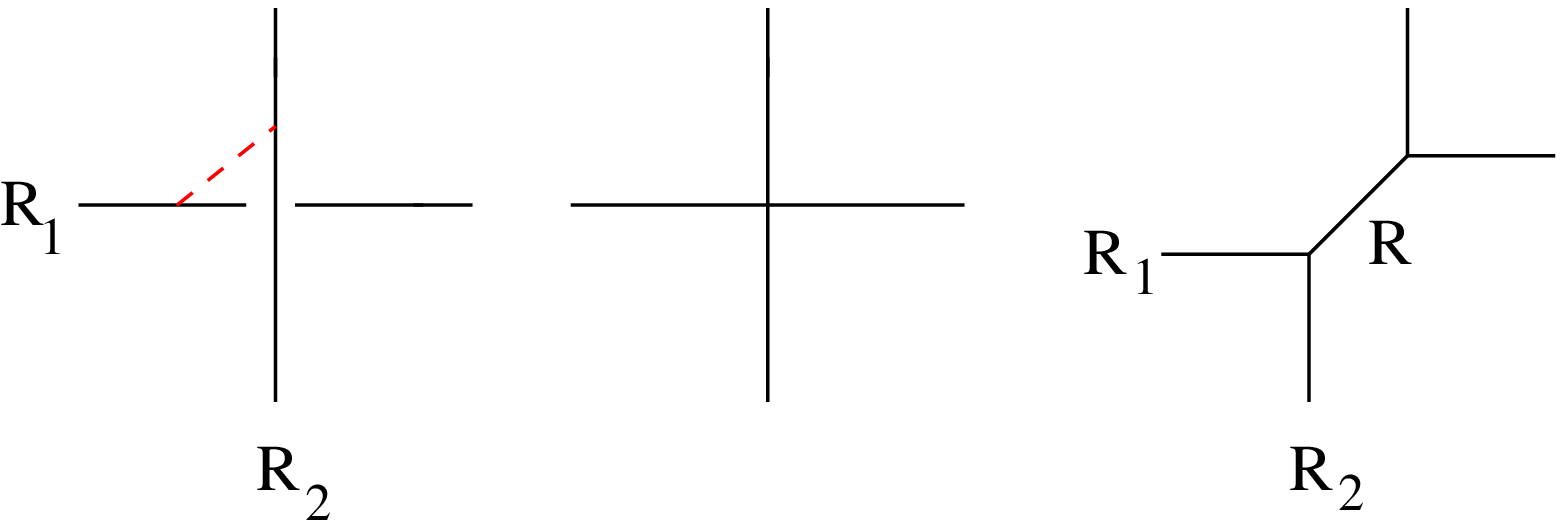}{A local transition.}
Assuming the existence of the three point vertex, the contribution to the partition function coming from the the diagram on the 
right (RHD) should be given by
\bea
Z_{R_1 R_2}^{\mbox{\tiny RHD}}=\sum_{R}Q^{l_{R}}\,V_{RR_{1}R_{2}}(q)\,{\cal W}_{R}(q)\,(-1)^{l_{R}}q^{-\frac{\kappa_{R}}{2}}\,,
\label{vertexeq}
\eea
where the factor of $(-1)^{l_{R}}q^{-\frac{\kappa_{R}}{2}}$ is 
due to the $\sl2z$ transformation, $S^{-1}T^{-1}$, which maps 
$\begin{pmatrix}1\cr 1\end{pmatrix}$ to $\begin{pmatrix}1\cr 0\end{pmatrix}$.

On the other hand, we know that the diagram on the left (LHD) is given by the CS expectation value of Wilson loops on a Hopf link with the two components in the representations $R_1$ and $R_2$,
\bea
Z_{R_1 R_2}^{\mbox{\tiny LHD}}=\lambda^{-\frac{l_{R_{1}}+l_{R_{2}}}{2}}W_{R_{1} R_{2}}\sum_{R}\,\lambda^{-l_{R}}{\cal W}_{R}^{2}(q)\,(-1)^{l_{R}}q^{-\frac{\kappa_{R}}{2}}\,.\,
\label{conifold}
\eea
In the above expression $\log \lambda$ is the K\"ahler parameter of the 
$\IP^{1}$, and $\lambda^{-\frac{l_{R_{1}}+l_{R_{2}}}{2}}$ is the renormalization factor discussed above. Equating (\ref{vertexeq}) and (\ref{conifold}) yields
\bea
\sum_{R}Q^{l_{R}}\,V_{RR_{1}R_{2}}(q)\,{\cal W}_{R}(q)\,(-1)^{l_{R}}q^{-\frac{\kappa_{R}}{2}}\,=\lambda^{-\frac{l_{R_{1}}+l_{R_{2}}}{2}}W_{R_{1} R_{2}}\sum_{R}\,\lambda^{-l_{R}}{\cal W}_{R}^{2}(q)\,(-1)^{l_{R}}q^{-\frac{\kappa_{R}}{2}}\,.
\eea 
We now set $Q=\lambda^{-1}$ and expand both sides in $\lambda^{-1}$ to obtain the following expressions for the three point vertex,
\bea \label{vertices}
V_{\cdot R_{1} R_{2}}&=&{\cal W}_{R_{1} R_{2}}\,,\\ \nn
V_{\tableau{1} R_{1} R_{2}}&=&{\cal W}_{\tableau{1}}{\cal W}_{R_{1} R_{2}}-
\frac{{\cal G}_{R_{1} R_{2}}(q)}{{\cal W}_{\tableau{1}}}\,,
\eea
where ${\cal G}_{R_{1} R_{2}}$ is the next to leading order coefficient in the expansion of $\lambda^{-\frac{l_{R_{1}}+l_{R_{2}}}{2}}W_{R_{1} R_{2}}$,
\bea
\lambda^{-\frac{l_{R_{1}}+l_{R_{2}}}{2}}W_{R_{1} R_{2}}={\cal W}_{R_{1} R_{2}}(q)+\lambda^{-1} {\cal G}_{R_{1} R_{2}}(q)+\ldots \,.
\eea
${\cal G}_{R_{1} R_{2}}$ can be determined easily from $W_{R_{1} R_{2}}$,
\bea \label{gs}
{\cal G}_{\cdot R}(q)&=&-{\cal W}_{R}\,f_{R}(q^{-1})\,,\\ \nn
{\cal G}_{\tableau{1} R}(q)&=&-{\cal W}_{R}\,{\cal W}_{\tableau{1}}-{\cal W}_{R \tableau{1}}\,f_{R}(q^{-1})\,,
\eea
where $f_{R}(q)=\sum_{i=1}^{d}\sum_{v=1}^{\mu_{i}}q^{v-i}$.
Thus from (\ref{vertices}) and (\ref{gs}), it follows that
\bea
V_{\cdot R \tableau{1}}(q)&=&{\cal W}_{R \tableau{1}}(q)\,,\\\nn
V_{\tableau{1} R \cdot}(q)&=&{\cal W}_{\tableau{1}}{\cal W}_{R}+\frac{{\cal W}_{R}}{{\cal W}_{\tableau{1}}}f_{R}(q^{-1})\\
&=&{\cal W}_{\tableau{1} R^{T}}q^{\kappa_{R}/2}\,,\\ \nn
V_{\tableau{1} R \tableau{1}}(q)&=&{\cal W}_{\tableau{1}}{\cal W}_{R \tableau{1}}+{\cal W}_{R}(1+f_{R}(q^{-1})\frac{{\cal W}_{R \tableau{1}}}{{\cal W}_{R}{\cal W}_{\tableau{1}}})\\
&=&{\cal W}_{R}\{1+\frac{{\cal W}_{\tableau{1} R}}{{\cal W}_{R}}\frac{{\cal W}_{\tableau{1} R^{T}}}{{\cal W}_{R^{T}}}\}\,,
\eea
where we have used the following identities,
\bea
\sum_{i=1}^{d}\sum_{v=1-i}^{\mu_{i}-i}q^{-v}&=&\frac{q}{(q-1)^{2}}\{\frac{{\cal W}_{\tableau{1} R^{T}}}{{\cal W}_{R^{T}}{\cal W}_{\tableau{1}}}-1\}\,,\\
{\cal W}_{R^{T}}(q)&=&{\cal W}_{R}(q)\,q^{-\kappa_{R}/2}\,,\label{transpose} \\ 
\frac{{\cal W}_{\tableau{1} R}}{{\cal W}_{\tableau{1}}}(q^{-1})&=&-\frac{{\cal W}_{\tableau{1} R}}{{\cal W}_{\tableau{1}}}(q)\,.
\eea
The expression for the vertices given above can be simplified and written
as follows
\bea
\frac{V_{\cdot R \tableau{1}}}{{\cal W}_{R}}&=&h_{R}(q),\\ \nn
\frac{V_{\tableau{1} R \cdot}}{{\cal W}_{R}}&=&h_{R^{T}}(q)\,,\\ \nn
\frac{V_{\tableau{1} R \tableau{1}}(q)}{{\cal W}_{R}}&=&1+h_{R}(q)\,h_{R^{T}}(q)\,,
\label{verticessimple}
\eea
where
\bea \label{definingh}
h_{R}(q):=\frac{{\cal
 W}_{\tableau{1} R}}{{\cal W}_{R}}\,={\cal W}_{\tableau{1}}+\frac{f_{R}(q)}{{\cal W}_{\tableau{1}}}\,.
\eea

\section{Calculating the partition function} \label{allorders}
Using geometric transitions to calculate closed string amplitudes along the lines introduced in \cite{AMV} computationally involves sums over all representations of $SU(N)$. Aborting the calculation at representations of a certain length, one obtains the Gopakumar-Vafa invariants $N^{g}_{(k,l,\ldots)}$ for all genera but only up to a restricted level in $k,l,\ldots$.

In \cite{firstpaper}, we presented a method which, for $A_1$ fibrations over $\IP^1$, yields the invariants to all orders in the fiber. This allows us to determine the generating functions
\begin{eqnarray}
f^{(k)}_g = \sum N^{g}_{(k,l)} Q_{F}^l  \;,
\end{eqnarray}
where the maximal level $k$ depends on the maximal length of representations we consider. In this section, we will expand the method to $A_{N-1}$ fibrations over $\IP^1$.

\subsection{$SU(2)$} \label{su2}
Let us first recall how we proceeded in the case of local Hirzebruch surfaces $\IF_m$, $m=0,1,2$. The relevant CS amplitude is given by \cite{firstpaper}
\begin{eqnarray}
Z_{CS}(Q_B,Q_F;q)&=&\sum_{R_{1,2,3,4}}Q_B^{l_{R_{1}}+l_{R_{3}}} Q_F^{m l_{R_{1}}+l_{R_{2}}+l_{R_{4}}} {\cal W}_{R_{1}R_{4}}(q)\,{\cal W}_{R_{4}R_{3}}(q){\cal W}_{R_{3}R_{2}}(q){\cal W}_{R_{2}R_{1}}(q) \nn \\
&& (-1)^{m(l_{R_1}+l_{R_3})}q^{\frac{m}{2}(\kappa_{R_1}-\kappa_{R_3})}\,.
\end{eqnarray}
To obtain the exact result to a given order in $Q_B$, we need to be able to 
perform the sum 
\begin{eqnarray} \label{kdefined}
K_{R_{1} R_{2}}(Q)=\sum_{R}Q^{l_{R}}{\cal W}_{R_{1}R}(q)\,{\cal W}_{RR_{2}}(q)\,.
\end{eqnarray}
In the case that $R_1$ and $R_2$ are trivial, this expression, 
$K_{\cdot \cdot}(Q)$, has a closed string interpretation: it is the partition function of $T^*(\IP^1)\times \IC$. This partition function was determined in \cite{AMV}. It is
\begin{eqnarray} 
K_{\cdot \cdot}(Q)=\mbox{Exp}\{\sum_{n=1}^{\infty}\frac{1}{n}{\cal W}^{2}_{\tableau{1}}(q^n)Q^{n}\}\,,\,\,\,\,\,{\cal W}_{\tableau{1}}(q)=\frac{1}{\sqrt{q}-\frac{1}{\sqrt{q}}}\,,
\end{eqnarray}
i.e. $N^g_{m}= -\delta_{g,0} \delta_{m,1}$. Inspired by this, we parametrize $K_{R_{1} R_{2}}$ for general $R_1$ and $R_2$ in the following way,
\begin{eqnarray} \label{kansatz}
K_{R_{1} R_{2}}(Q)=
{\cal W}_{R_{1}}(q){\cal W}_{R_{2}}(q)\mbox{Exp}\{\sum_{n=1}^{\infty}{\tilde f}_{R_{1}R_{2}}^{n}(q)Q^{n}\}\,.
\end{eqnarray}
The coefficient of the exponential is dictated by comparison of the $0$th order term in $Q$ between (\ref{kdefined}) and (\ref{kansatz}). So far, nothing is gained, since we must determine the unknown functions ${\tilde f}_n$ for all $n$. The crucial assumption we now make is that all ${\tilde f}_n$ can be determined from ${\tilde f}_1$ via
\begin{eqnarray}
{\tilde f}_{R_{1}R_{2}}^{n}(q)=\frac{{\tilde f}^1_{R_{1}R_{2}}(q^{n})}{n}\,.
\end{eqnarray}
This form for the coefficients of the multicovering is not that surprising 
since the term in the exponential in this case is a refinement of the open 
string amplitude which is conjectured \cite{OV} to have multicovering 
contribution with coefficients satisfying the above equation.
Once we conjecture this form for ${\tilde f}^n_{R_{1}R_{2}}$, determining $K_{R_{1}R_{2}}(Q)$ is a matter of determining ${\tilde f}_{R_{1}R_{2}}^1$. This can be achieved by expanding (\ref{kdefined}) and (\ref{kansatz}) to first order in $Q$. A simple calculation yields
\bea
{\tilde f}^{1}_{R_{1}R_{2}}(q)&=&\frac{{\cal W}_{R_{1} \tableau{1}}}{{\cal W}_{R_{1}}}
\frac{{\cal W}_{\tableau{1} R_{2}}}{{\cal W}_{R_{2}}}\,\\ \nn
&=&
\frac{q}{(q-1)^{2}}\{1+(q-1)\sum_{j=1}^{d_{1}}(q^{\mu^1_{j}-j}-q^{-j})\}\{1+(q-1)\sum_{j=1}^{d_{2}}(q^{\mu^2_{j}-j}-q^{-j})\}\,.
\eea
In the following, it will often be convenient to consider the quantity
\begin{eqnarray} \label{khat}
{\widehat K}_{R_1 R_2} &=& \frac{K_{R_{1} R_{2}}(Q)}{K_{\cdot \cdot}(Q)} \\
&=&{\cal W}_{R_{1}}{\cal W}_{R_{2}}\,\mbox{Exp}\{\sum_{n}\frac{f_{R_{1} R_{2}}(q^{n})}{n}Q^{n}\}\,.
\end{eqnarray}
The $f_{R_{1} R_{2}}$ are related in a simple way to the ${\tilde f}_{R_{1} R_{2}}$ and the $h_R$ defined in (\ref{definingh}),
\bea  \label{ffh}
f_{R_{1} R_{2}}(q)+{\cal W}_{\tableau{1}}^2(q) &=& \tilde{f}^{1}_{R_{1}R_{2}}(q) \\
&=& h_{R_{1}}(q)h_{R_{2}}(q) \,.
\label{deff}
\eea
The coefficients $f_{R_{1} R_{2}}$ were introduced in \cite{firstpaper}.
They can be written as a finite sum in powers of $q$ and $q^{-1}$,
\begin{eqnarray} \label{fexp}
f_{R_{1} R_{2}}(q)=\sum_{k}C_{k}(R_{1},R_{2})q^{k} \,,
\end{eqnarray}
such that the expansion coefficients have the following properties,
\bea
f_{R_{1} R_{2}}(1)&=&\sum_{k}C_{k}(R_{1},R_{2})=l_{R_{1}}+l_{R_{2}}\,,\\\nn
\frac{df_{R_{1} R_{2}}(q)}{dq}|_{q=1}&=&\sum_{k}k C_{k}(R_{1},R_{2})=\frac{\kappa_{R_{1}}+\kappa_{R_{2}}}{2}\,.
\eea

Substituting the expansion (\ref{fexp}) into (\ref{khat}), we obtain 
\bea\nn
\frac{K_{R_{1}R_{2}}(Q)}{K_{\cdot \cdot}(Q)}&=&{\cal W}_{R_{1}}{\cal W}_{R_{2}}\,\prod_{k}(1-q^{k}Q)^{-C_{k}(R_{1},R_{2})}\,.
\label{summ}
\eea
Thus the partition function of local $\IF_{m}$ is given by
\bea 
Z(Q_{B},Q_{F})&=&K_{\cdot \cdot}^{2}(Q_{F})\sum_{R_{1,2}} (-1)^{m(l_{R_1}+l_{R_2})}q^{\frac{m}{2}(\kappa_{R_1}-\kappa_{R_2})}  Q_{B}^{l_{R_{1}}+l_{R_{2}}} Q_{F}^{m l_{R_1}} \frac{{\cal W}^{2}_{R_{1}}{\cal W}^{2}_{R_{2}}}{\prod_{k}(1-q^{k}Q_{F})^{2C_{k}(R_{1},R_{2})}} \nn \\
&=& K_{\cdot \cdot}^{2}(Q_{F})\sum_{R_{1,2}}  Q_{B}^{l_{R_{1}}+l_{R_{2}}} Q_{F}^{m l_{R_1}} K_{R_{1}R_{2}}(Q_F) K^{(m)}_{R_{1}R_{2}}(Q_F)\,,  \label{psu2}
\eea
where we have defined $K^{(m)}_{R_{1}R_{2}}(Q)$ as
\begin{eqnarray}
K^{(m)}_{R_{1}R_{2}}(Q)= (-1)^{m(l_{R_1}+l_{R_2})}q^{\frac{m}{2}(\kappa_{R_1}-\kappa_{R_2})} K_{R_{1}R_{2}}(Q) \,.
\end{eqnarray}
This splitting of $Z(Q_{B},Q_{F})$ into contributions from $K$ and $K^{(m)}$ can be depicted as in \figref{splitting}. 
\onefigure{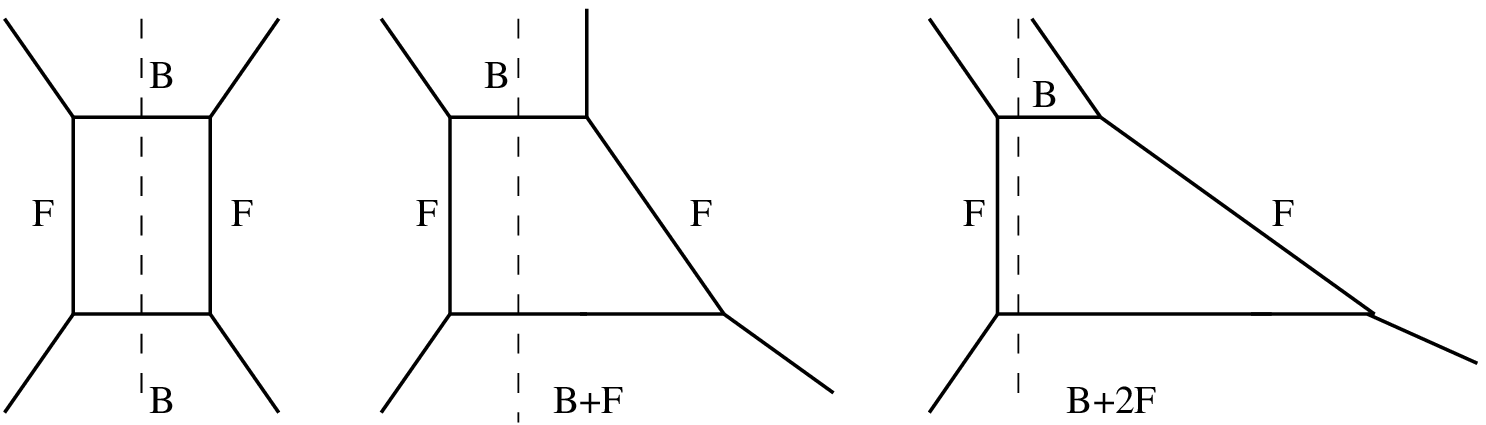}{Splitting local $\IF_m$ into $K^{(m)}_{R_i R_j}$ contributions.}

\subsection{$SU(3)$}
We now consider the case of $SU(3)$ geometries and generalize this to 
$SU(N)$ in the next subsection. There are four inequivalent geometries giving pure $SU(3)$ 
gauge theory via geometric engineering. The web diagram corresponding to 
these geometries are shown in \figref{su3}. We discuss how these diagrams come about in the appendix.
\onefigure{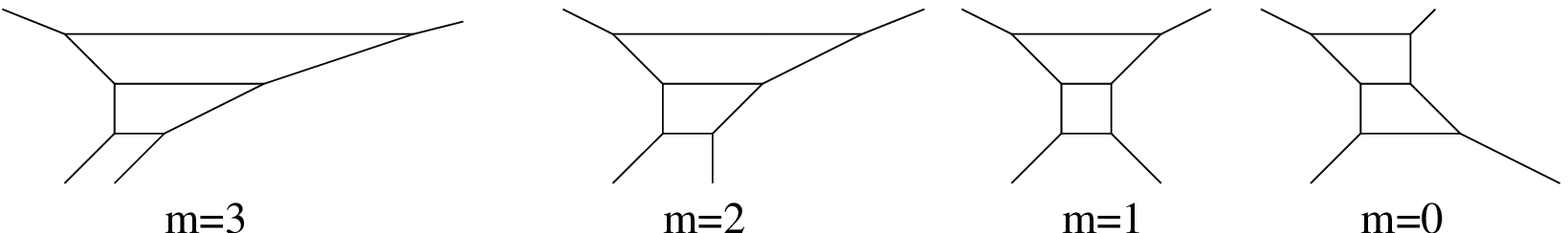}{a) m=3 with $\IF_{2}$ and $\IF_{4}$, b) m=2  with $\IF_{1}$ and $\IF_{3}$   c) m=1 with $\IF_{0}$ and $\IF_{2}$ d) m=0 with $\IF_{1}$ and $\IF_{1}$.}

\begin{figure}[h]
\begin{center}
\leavevmode\epsfbox{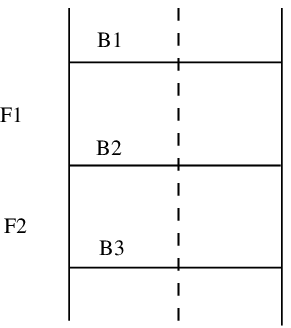}
\end{center} 
\caption{\small Ladder structure of web diagrams.}\label{ladder}
\end{figure}

Ignoring the angles, these web diagrams all have the ladder structure depicted in \figref{ladder}.

\psfrag{B1}{$B_1$}
\psfrag{B2}{$B_2$}
\psfrag{B3}{$B_3$}
\psfrag{F1}{$F_1$}
\psfrag{F2}{$F_2$}

The K\"ahler parameters $T_{b_{i}}$ are related to the K\"ahler
parameters of the base $B$ and fibers $F_{1,2}$ (recall that $Q_{c}=e^{-T_c}$) as follows,
\bea
Q_{b_{1}}&=&Q_{B}Q_{F_{1}}^{m+1}Q_{F_{2}}^{(m-1)(1-\delta_{m,0})}\,,\\ \nn
Q_{b {2}}&=&Q_{B}Q_{F_{2}}^{(m-1)(1-\delta_{m,0})}\,,\\ \nn
Q_{b_{3}}&=&Q_{B}Q_{F_{2}}^{\delta_{m,0}}\,.
\eea
In analogy to the $SU(2)$ case, we split the ladder across its rungs (dashed line in \figref{ladder}) and define for each half
\bea \label{khat3}
K^{(m)}_{R_{1}R_{2}R_{3}}(Q_{F_{1}},Q_{F_{2}})=M^{(m)}(q,R_{i})\sum_{S_{1,2}}
{\cal W}_{R_{1} S_{1}}(q)\,Q_{F_{1}}^{l_{S_{1}}}\,V_{S_{1} R_{2} S_{2}}(q)\,
Q_{F_{2}}^{l_{S_{2}}}\,{\cal W}_{S_{2} R_{3}}(q)\,,
\label{pfsu3}
\eea
where $M^{(m)}(q,R_{i})=q^{\sum_{i=1}^{3}\alpha_{i}(m)\frac{\kappa_{R_{i}}}{2}}\,(-1)^{\sum_{i=1}^{3}\alpha_{i}(m)l_{R_{i}}}$ is a framing factor.
The $Q_{F_1,F_2}$ independent term in the above expression is given by $K^{(m)}_{R_{1}R_{2}R_{3}}(Q_{F_1}=0,Q_{F_2}=0)=M^{(m)}(q,R_{i}){\cal W}_{R_{1}}{\cal W}_{R_{2}}{\cal W}_{R_{3}}$. We parametrize the $Q_{F_1,F_2}$ dependent pieces as follows  
\bea\label{expa}
K^{(m)}_{R_{1}R_{2}R_{3}}(Q_{F_1}Q_{F_2})&=&M^{(m)}(q,R_{i})\,{\cal W}_{R_{1}}\,{\cal W}_{R_{2}}\,{\cal W}_{R_{3}}\\ \nn
&&\mbox{Exp}\{\sum_{n=1}^{\infty}A^{(n)}_{R_{1}R_{2}}(q)Q_{F_{1}}^{n}
+\sum_{n=1}^{\infty}A^{(n)}_{R_{2}R_{3}}(q)Q_{F_{2}}^{n}+\sum_{n=1}^{\infty}A^{(n)}_{R_{1}R_{2}R_{3}}(q)(Q_{F_{1}}Q_{F_{2}})^{n}\}\,.
\eea
The three sums in the exponential are to take account of the holomorphic curves in the open string geometry and their multicovers running between the upper two, the lower two, and the upper and the lower rung of the ladder. Since this geometry, ignoring the representations on the external legs, is resolved $A_{2} \times \IC$, holomorphic cycles are in one to one correspondence with positive roots of $SU(3)$, i.e., there are three holomorphic cycles $F_{1},F_{2}$ and $F_{1}+F_{2}$. Now, we again make 
an assumption about the coefficient in the multicovering expansion,
\bea
A^{(n)}_{R_{1}R_{2}}(q)&=&\frac{A_{R_{1}R_{2}}(q^{n})}{n}\,,\,\,A^{(n)}_{R_{2}R_{3}}(q)=\frac{A_{R_{2}R_{3}}(q^{n})}{n}\,,\\\nn
A^{(n)}_{R_{1}R_{2}R_{3}}(q)&=&\frac{A_{R_{1}R_{2}R_{3}}(q^{n})}{n}\,.
\eea
Equating the coefficients  of $Q_1$, $Q_2$, and $Q_1 Q_2$ respectively in (\ref{pfsu3}) and (\ref{expa}) yields
\begin{eqnarray}\nn
A_{R_{1}R_{2}}(q) &=& \frac{{\cal W}_{\tableau{1} R_1} V_{\tableau{1} R_2 \cdot}}{{\cal W}_{R_1}{\cal W}_{R_2}}\\ \nn
 &=& h_{R_{1}}(q)h_{R_{2}^{T}}(q)\,,\\ \nn
A_{R_{2}R_{3}}(q)&=&\frac{V_{\cdot R_2 \tableau{1}}{\cal W}_{\tableau{1} R_3}}{{\cal W}_{R_2}{\cal W}_{R_3}} \\\nn
&=&h_{R_{2}}h_{R_{3}}\,,\\ \nn
A_{R_{1}R_{2}R_{3}}&=&\frac{{\cal W}_{R_1 \tableau{1}} V_{\tableau{1} R_2 \tableau{1}}{\cal W}_{\tableau{1} R_3}}{{\cal W}_{R_1}{\cal W}_{R_2}{\cal W}_{R_3}}-
A_{R_{1}R_{2}}A_{R_{2}R_{3}}\\
&=&h_{R_{1}}h_{R_{3}}\,.
\end{eqnarray}
By (\ref{ffh}), we hence obtain
\bea\nn
\widehat{K}^{(m)}_{R_{1}R_{2}R_{3}}(Q_{F_{1}},Q_{F_{2}})&:=&\frac{K^{(m)}_{R_{1}R_{2}R_{3}}(Q_{F_{1}},Q_{F_{2}})}{K_{00}(Q_{F_{1}})K_{00}(Q_{F_{2}})K_{00}(Q_{F_{1}}Q_{F_{2}})}  \\\nn
&=&\frac{M^{(m)}(q,R_{i}){\cal W}_{R_{1}}\,{\cal W}_{R_{2}}\,{\cal W}_{R_{3}}}
{\prod_{k}(1-q^{k}Q_{F_{1}})^{C_{k}(R_{1},R_{2}^{T})}(1-q^{k}Q_{F_{2}})^{C_{k}(R_{2},R_{3})}(1-q^{k}Q_{F_{1}}Q_{F_{2}})^{C_{k}(R_{1},R_{3})}}\,. 
\eea
Define
\bea
\widehat{Z}^{(m)}(Q_{F_{1}},Q_{F_{2}}):=\frac{Z^{(m)}(Q_{F_{1}},Q_{F_{2}})}{K_{00}(Q_{F_{1}})^2 K_{00}(Q_{F_{2}})^2 K_{00}(Q_{F_{1}}Q_{F_{2}})^2}\,,
\eea
then
\bea\nn
\widehat{Z}^{(m)}(Q_{1},Q_{2})&=&\sum_{R_{1,2,3}}e^{-T_{b_{1}}l_{R_{1}}-T_{b_{2}}l_{R_{2}}-T_{b_{3}}l_{R_{3}}}\widehat{K}^{(m_{1})}_{R_{1}R_{2}R_{3}}(Q_{1},Q_{2})\,\widehat{K}^{(m_{2})}_{R_{1}R_{2}R_{3}}(Q_{1},Q_{2})\\ \nn
&=&
\sum_{R_{1,2,3}}Q_{b_{1}}^{l_{R_{1}}}\,Q_{b_{2}}^{l_{R_{2}}}\,Q_{b_{3}}^{l_{R_{3}}}\,M^{(m_{1})}(q,R_{i})M^{(m_{2})}(q,R_{i})\\ \nn
&&\frac{{\cal W}_{R_{1}}^{2}\,{\cal W}_{R_{2}}^{2}\,{\cal W}_{R_{3}}^{2}}{\prod_{k}(1-q^{k}Q_{1})^{2C_{k}(R_{1},R_{2}^{T})}(1-q^{k}Q_{2})^{2C_{k}(R_{2},R_{3})}(1-q^{k}Q_{1}Q_{2})^{2C_{k}(R_{1},R_{3})}}\\ \nn
&=&\sum_{R_{1,2,3}}\,Q_{B}^{l_{1}+l_{2}+l_{3}}Q_{F_{1}}^{(m+1)l_{1}}Q_{F_{2}}^{(m-1)
(1-\delta_{m,0})(l_{1}+l_{2})+\delta_{m,0}l_{3}}\\ \label{psu3}
&& M(q,R_{i}) \frac{{\cal W}_{R_{1}}^{2}\,{\cal W}_{R_{2}}^{2}\,{\cal W}_{R_{3}}^{2}}{\prod_{1\leq i<j\leq 3}\prod_{k}(1-q^{k}Q_{ij})^{2C_{k}(R_{i},R^{T}_{j})}}\,,
\eea
where $Q_{12}=Q_{F_{1}},Q_{23}=Q_{F_{2}}$ and $Q_{13}=Q_{F_{1}}Q_{F_{2}}$. Also in the third line above we have changed $R_{3}$ to $R_{3}^{T}$ and 
used (\ref{transpose}), so that
\begin{eqnarray}
M(q,R_{i})&=&M^{(m_{1})}(q,R_{i})M^{(m_{2})}(q,R_{i})q^{-\kappa_3} \\ \nn
&=& (-1)^{\alpha l_1+\beta l_2 + \gamma l_3}q^{\frac{1}{2}(\alpha \kappa_1 + \beta \kappa_2 + \gamma \kappa_3)}\,.
\end{eqnarray}
To determine the framing factors $\alpha$, $\beta$, $\gamma$, which of course depend on $m$, we take $Q_{F_1} \rightarrow 0$, $Q_{F_2} \rightarrow 0$ respectively and thus reduce to the local Hirzebruch geometries we studied in \cite{firstpaper} and reviewed in the previous section. To compare the limit of (\ref{psu3}) to the $SU(2)$ partition function, (\ref{psu2}), we use (\ref{transpose}) to rewrite the $SU(3)$ partition function in terms of non-transposed representations. For the geometry containing $\IF_k$, $\IF_{l}$, $(k,l)\neq(1,1)$, we obtain $\alpha = k$, $\beta=k-2$ from the $Q_{F_2} \rightarrow 0$ limit, and $\beta = l$, $\gamma=l-2$ from the $Q_{F_1} \rightarrow 0$ limit.\footnote{In \cite{firstpaper}, we consider the local Hirzebruch surfaces $\IF_k$ for $k=0,1,2$, since the canonical line bundle over higher Hirzebruch surfaces contains additional compact divisors. Here, we match the $Q_{F_1,F_2} \rightarrow 0$ limit to (\ref{psu2}) for arbitrary $k$, and show that this allows a consistent choice of framing for all $SU(3)$ geometries.}  Since $k$ and $l$ are related via $l=k-2$, we can choose a consistent framing for the full geometry. For the case $(k,l)=(1,1)$, the $Q_{F_1} \rightarrow 0$ limit yields $\beta = -l$, $\gamma=-l-2$, again consistent with a choice of framing for the full geometry. In terms of the integer $m$ used to label the $SU(3)$ geometries in \figref{su3}, the framing coefficients are $\alpha=m+1$, $\beta=m-1$, $\gamma=m-3$.

\subsubsection{Gopakumar-Vafa invariants}
We can evaluate (\ref{psu3}) to obtain generating functions for Gopakumar-Vafa invariants,
\begin{eqnarray}
f_g^{(n)} (x,y)&=& \sum_{k,l}(-1)^{g-1} N_{(n,k,l)}^g x^k y^l \,.
\eea
We consider the $m=1$ and $m=3$ case. The latter was also considered in \cite{AKMV}. 
For $m=3$ and $n=1$, we obtain
\begin{eqnarray}
f_g^{(1)} (x,y)&=& \delta_{g,0} \left( \frac{y^2}{(1-x)^2 (1-y)^2}+\frac{1}{(1-y)^2 (1-x y)^2}+\frac{x^4 y^2}{(1-x)^2 (1-x y)^2} \right) \,,
\end{eqnarray}
which agrees with \cite{AKMV} upon expansion in $x$ and $y$. For $n=2$, the expression for the generating function is too long to reproduce here. It has the form  
\bea
f^{(2)}_{g}(x,y)&=&\frac{P_{2,g}(x,y)}{(1-x)^{2g+6}(1-y)^{2g+6}(1-xy)^{2g+6}(1+x)^{2}(1+y)^{2}(1+xy)^{2}}\,,
\end{eqnarray}
with $P_{2,0}$ a polynomial of order $13$ in $x$, $16$ in $y$, $P_{2,1}$ a polynomial of order $16$ in $x$, $20$ in $y$, etc.
We can expand these expressions out to low order in $x$ and $y$ to get
\begin{eqnarray}
f^{(2)}_{0}(x,y) &=& 6 y^3 + 32 y^4 + 110 y^5 + (10 y^3 +70 y^4 + 270 y^5)x +(12 y^3 + 96 y^4+416 y^5 )x^2 \nn \\ 
& &+(12 y^3+ 110 y^4 + 518 y^5)x^3 +(y^3+ 112 y^4 + 576 y^5)x^4\nn \\
& &+(14 y^3+ 126 y^4 + 630 y^5)x^5 +\ldots \nn \\
f^{(2)}_{1}(x,y) &=& 9 y^4+68 y^5 + (16 y^4 + 144 y^5) x + (21 y^4 + 204 y^5) x^2 + (24 y^4 + 248 y^5) x^3 \nn \\
& &+(25y^4+276y^5)x^4  + (24y^4+288y^5)x^5 \ldots \nn\\
f^{(2)}_{2}(x,y) &=& 12 y^5 + 22 y^5 x + 30 y^5 x^2 + 36 y^5 x^3 + 40 y^5 x^4 +42 y^5 x^5 + \ldots \,.\nn
\end{eqnarray}
For $m=1$, we obtain
\begin{eqnarray}
f_g^{(1)} (x,y)&=& \delta_{g,0} \left( \frac{1}{(1-x)^2 (1-y)^2}+\frac{1}{(1-y)^2 (1-x y)^2}+\frac{x^2}{(1-x)^2 (1-x y)^2} \right) \,,
\end{eqnarray}
and
\begin{eqnarray}
f^{(2)}_{0}(x,y) &=& 6 y + 32 y^2 + 110 y^3 +288y^4+644y^5 + (10 y +70 y^2 + 270 y^3+770y^4+1820y^5)x \nn \\
& & +(12 y + 96 y^2+416 y^3+1280y^4 +3204 y^5 )x^2 \nn \\
 & &+(6 + 30y +140y^2+560y^3+1764 y^4 + 4576 y^5)x^3 \nn \\
& & +(32+98y+288y^2+ 840y^3+ 2368 y^4 + 6020 y^5)x^4\nn \\
& &+(110+306y+672y^2+1540 y^3+ 3528 y^4 + 8064 y^5)x^5 +\ldots \nn \\
f^{(2)}_{1}(x,y) &=& 9 y^2+68 y^3 +300y^4+988y^5 + (16 y^2 + 144 y^3+704y^4+2496y^5) x \nn \\ 
& &+ (21 y^2 + 204 y^3+1073y^4+4032y^5) x^2 + (24 y^2 + 248 y^3 + 1368 y^4 +5368 y^5) x^3\nn \\ 
& & + (9+16y+57y^2+324y^3+1653y^4+6528y^5)x^4 \nn \\ 
& &+(68+144y+252y^2+668y^3+2268y^4+7956y^5)x^5 \ldots \nn\\
f^{(2)}_{2}(x,y) &=& 12 y^3 +116 y^4 +628 y^5+ (22 y^3 +242y^4+1430y^5)x + (30 y^3+348y^4+2168y^5) x^2 \nn \\ 
& & + (36 y^3 +434 y^4 + 2794 y^5) x^3 + (40 y^3 + 500 y^4 +3308 y^5)x^4\nn \\ 
& & +(12+22y+30y^2+92y^3+616y^4+3800y^5) x^5 + \ldots \,. \nn
\end{eqnarray}
\subsection{$SU(N)$}
In this section we generalize the calculation of the previous
section to the case of geometries giving rise to $SU(N)$ gauge theory
via geometric engineering. 

\begin{figure}[h]
\begin{flushleft}\leavevmode\epsfbox{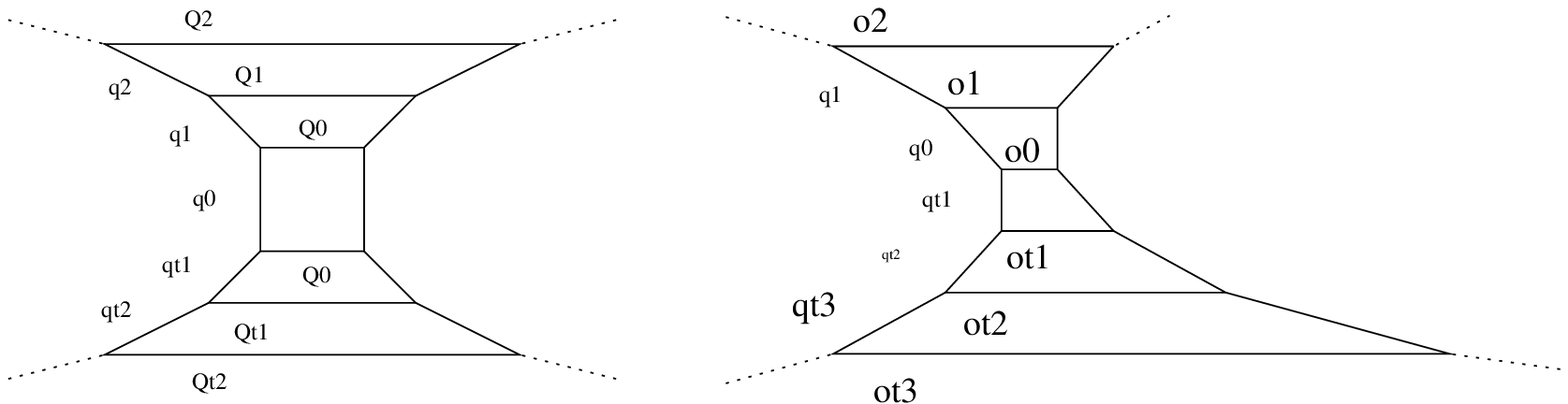}\end{flushleft} 
\caption{\small Identifying the K\"ahler classes of curves, for geometries containing even Hirzebruch surfaces (left), and odd (right).} \label{pattern}
\end{figure}

Consider the half-web shown in \figref{halfweb} below. Two such webs joined together give rise to the web diagrams for $SU(N)$ geometries depicted in \figref{sunweb}. 
\psfrag{K}{$K_{R_1 \cdots R_N}^{(m)}(Q_{F_{1,\cdots,N-1}})=$}
\onefigure{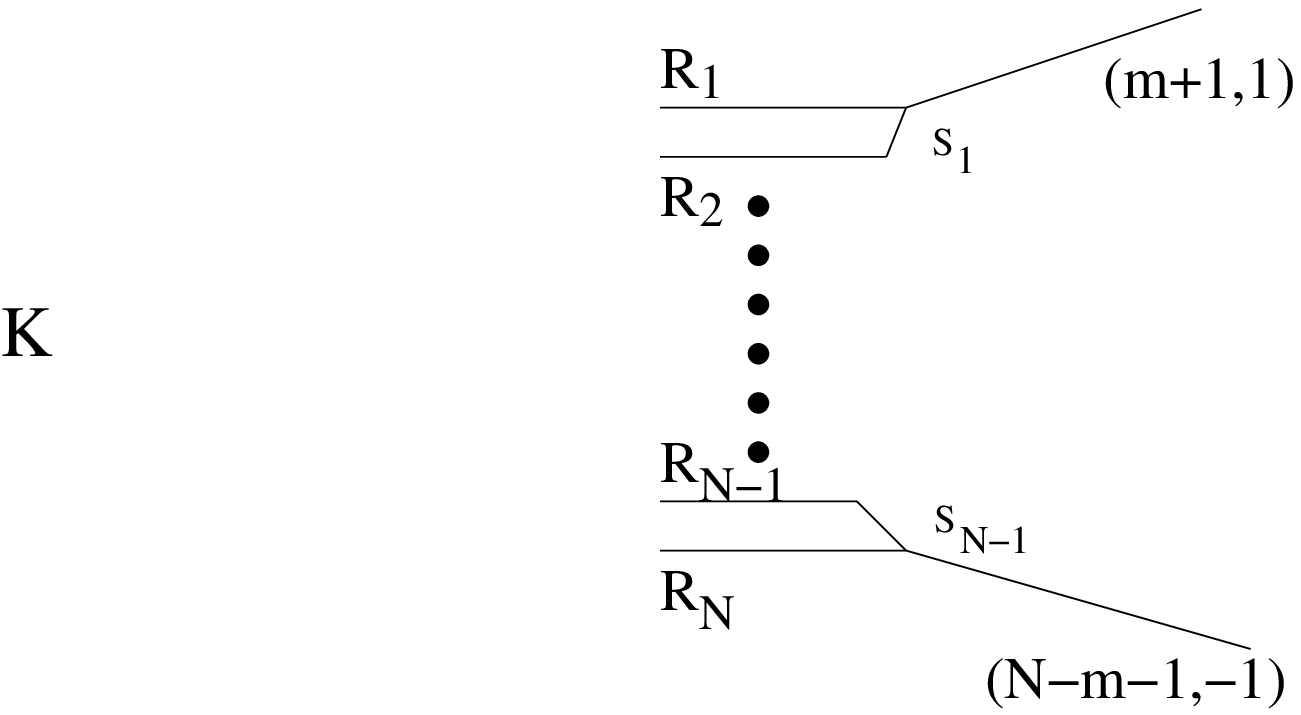}{The diagrammatic representation of $K^{(m)}_{R_{1}\cdots R_{N}}$. The tupels in parentheses denote the slopes of the corresponding lines.}
As in the case of $SU(3)$ geometries discussed in the last section, the partition function associated with such a half-web is given by
\bea
K^{(m)}_{R_{1}\cdots R_{N}}(Q_{F_{1,\cdots,N-1}})&=&M^{(m)}(q,R_{i})\sum_{S_{1,\cdots,N-1}}
{\cal W}_{R_{1}S_{1}}(q)Q_{1}^{l_{S_{1}}}V_{S_{1}R_{2}S_{2}}(q)\cdots \\ \nn
&&V_{S_{N-2}R_{N-1}S_{N-1}}(q)Q_{N-1}^{l_{S_{N-1}}}{\cal W}_{S_{N-1}R_{N}}(q)\,q^{\sum_{i=1}^{N-1}\beta_{i}\kappa_{S_{i}}}\,(-1)^{\sum_{i=1}^{N-1}\beta_{i}l_{S_{i}}}\,,
\label{pfsum}
\eea
where $M^{(m)}(q,R_{i})=q^{\sum_{i=1}^{N}\frac{\alpha_{i}}{2}\kappa_{R_{i}}}\,(-1)^{\sum_{i=1}^{N}\alpha_{i}l_{R_{i}}}$. Since the geometry described by the half-web (i.e. ignoring the representations on the external legs) is resolved $A_{N-1}\times \IC$, we know that the holomorphic cycles are in one to one correspondence with the positive roots of $SU(N)$ such that the cycles $F_{i}$ correspond to the simple roots. The above partition function can be
written as a sum over all holomorphic curves in the geometry and their
multicovering,
\bea\nn
K^{(m)}_{R_{1}\cdots R_{N}}(Q_{1,\cdots,N-1})&=&M^{(m)}(q,R_{i})\,{\cal W}_{R_{1}}(q)
\cdots {\cal W}_{R_{N}}(q)\,\\ \nn
&&\mbox{Exp}\{\sum_{i=1}^{N-1}\sum_{r=0}^{N-1-i}\sum_{n=1}^{\infty}
A^{(n)}_{R_{i}\cdots R_{i+r+1}}(q)(Q_{i}Q_{i+1}\cdots Q_{i+r})^{n}\}\,,\\\nn
&=&M^{(m)}(q,R_{i})\,{\cal W}_{R_{1}}(q)
\cdots {\cal W}_{R_{N}}(q)\,\\ 
&&\mbox{Exp}\{\sum_{i=1}^{N-1}\sum_{r=0}^{N-1-i}\sum_{n=1}^{\infty}
\frac{A_{R_{i}\cdots R_{i+r+1}}(q^{n})}{n}(Q_{i}Q_{i+1}\cdots Q_{i+r})^{n}\}\,,
\eea
where, as before, we have assumed 
\bea
A^{(n)}_{R_{i}\cdots R_{i+r+1}}(q)=\frac{A_{R_{i}\cdots R_{i+r+1}}(q^{n})}{n}\,.
\eea
The functions $A_{R_{i}\cdots R_{j}}(q)$ can be easily determined from (\ref{pfsum}) and (\ref{verticessimple}),
\bea
A_{R_{i}\cdots R_{j}}&=&h_{R_{i}}(q)\,h_{R_{j}^{T}}(q)\,,\,\,j\neq N\,,\\ \nn
A_{R_{i}\cdots R_{N}}&=&h_{R_{i}}(q)\,h_{R_{N}}\,,\,i=1,\cdots,N-1\,.
\eea
Define
\bea
\widehat{K}^{(m)}_{R_{1}\cdots R_{N}}(Q_{1},\cdots,Q_{N-1})=\frac{K^{(m)}_{R_{1}\cdots R_{N}}}{\prod_{1\leq i<j\leq N}K_{00}(Q_{i}\cdots Q_{j-1})}\,,
\eea
then using (\ref{summ}) we get
\bea\nn
\widehat{K}^{(m)}_{R_{1}\cdots R_{N}}(Q_{1},\cdots,Q_{N-1})&=&M(q,R_{i})\prod_{i=1}^{N}{\cal W}_{R_{i}}^{2}\,\prod_{1\leq i<j<N-1}(1-q^{k}Q_{i}\cdots Q_{j-1})^{-C_{k}(R_{i},R_{j})}\\ \nn
&&\prod_{i=1}^{N}(1-q^{k}Q_{i}\cdots Q_{N-1})^{-C_{k}(R_{i},R_{N})}\,.
\eea
The partition function is given by
\bea
{\widehat Z}^{(m)} &=& \sum_{R_{1,\cdots,N}}(\prod_{i=1}^{N}Q_{b_{i}}^{l_{R_{i}}})\widehat{K}^{(N-2)}_{R_{1}\cdots R_{N}}(Q_{1},\cdots,Q_{N-1})   \widehat{K}^{(m)}_{R_{1}\cdots R_{N}}(Q_{1},\cdots,Q_{N-1}) \nn \\
&=&\sum_{R_{1,\cdots,N}}(\prod_{i=1}^{N}Q_{b_{i}}^{l_{R_{i}}})
M(q,R_{i})\frac{\prod_{i=1}^{N}{\cal W}_{R_{i}}(q)^{2}}{\prod_{1\leq i<j\leq N}\prod_{k}(1-q^{k}Q_{i}\cdots Q_{j-1})^{2C_{k}(R_{i},R_{j}^{T})}}\,.
\eea
In writing the above expression we have changed $R_{N}$ to $R_{N}^{T}$ and
absorbed a factor of $q^{-\kappa_{N}}$ into $M^{(m)}(q,R_{i})$.
Studying the limits $Q_{F_i}\rightarrow 0$ as in the previous section, we can determine the framing factor to be
\bea
M^{(m)}(q,R_{i})=(-1)^{(N+m)(l_{1}+\cdots+ l_{N})}\,q^{\frac{1}{2}\sum_{i=1}^{N}(N+m-2i)\kappa_{i}}\,.
\label{sunframing}
\eea
Recall that when $N+m$ is even (odd), the corresponding geometry contains 
Hirzebruch surfaces $\ff_{2r}$ ($\ff_{2r+1}$). The K\"ahler parameters
$T_{b_{i}}$ are related to the K\"ahler parameter of the
base $T_{B}$ and of the fibers $T_{F_{1,\cdots,N-1}}$ via,

for $N+m=2r+1$,
\begin{eqnarray}
T_{b_{r+1}}&=&T_{B}\,,\\ \nn
T_{b_{r+1-i}}&=&T_{B}+\sum_{j=1}^{i}(2j-1)T_{F_{r+1-j}}\,,\,\,i=1,\cdots,r\,,\\ \nn
T_{b_{r+1+i}}&=&T_{B}+\sum_{j=1}^{i}(2j-1)T_{F_{r+j}}\,,\,\,i=1,\cdots,N-r-1\,,\\ \nn
\end{eqnarray}
and for $N+m=2r$,
\begin{eqnarray}
T_{b_{r}}&=&T_{b_{r+1}}=T_{B} \,,\\\nn 
T_{b_{r-i}}&=&T_{B}+\sum_{j=1}^{i}2j\,T_{F_{r-j}}\,,i=1,\cdots,r-1\\ \nn
T_{b_{r+1+i}}&=&T_{B}+\sum_{j=1}^{i}2j\,T_{F_{r+j}}\,,i=1,\cdots,N-r-1\,,
\label{k2}
\end{eqnarray}
where $T_{b_{i}}$ is to be set to $0$ if $i>N$ or $i<1$.
The above relation between the K\"ahler parameters is depicted in \figref{pattern}. It can be determined from the fact
that the divisors which appear in the local geometry are
\bea
\{\ff_{m+2-N},\ff_{m+4-N},\cdots,
\ff_{m+2r-N},\cdots,\ff_{m+N-2}\}\,,
\eea
where $\IF_{-n}$ for $n>0$ is to indicate that the corresponding subdiagram within the web diagram for the $SU(N)$ geometry occurs upside down as compared to the subdiagram for $\IF_{n}$.

\psfrag{q0}{$F_r$}
\psfrag{q1}{$F_{r-1}$}
\psfrag{q2}{$F_{r-2}$}
\psfrag{qt1}{$F_{r+1}$}
\psfrag{qt2}{$F_{r+2}$}
\psfrag{qt3}{$F_{r+3}$}
\psfrag{Q0}{$B$}
\psfrag{Q1}{$B+2 F_{r-1}$}
\psfrag{Q2}{$B+2 F_{r-1} + 4 F_{r-2}$}
\psfrag{Qt1}{$B +2F_{r+1}$}
\psfrag{Qt2}{$B +2F_{r+1}+4 F_{r+2}$}
\psfrag{o0}{$B$}
\psfrag{o1}{$B+ F_{r}$}
\psfrag{o2}{$B+ F_{r}+ 3 F_{r-1}$}
\psfrag{ot1}{$B + F_{r+1}$}
\psfrag{ot2}{$B+ F_{r+1}+3 F_{r+2}$}
\psfrag{ot3}{$B+ F_{r+1}+3 F_{r+2} +5 F_{r+3}$}

From (\ref{k2}) we can easily determine $\prod_{i}Q_{b_{i}}^{l_{i}}$ to be
\bea\nn
\prod_{i=1}^{N}Q_{b_{i}}^{l_{R_{i}}}
= 
\begin{cases}
Q_{B}^{\sum_{i=1}^{N}l_{R_{i}}}
\,\prod_{i=1}^{\frac{N+m-1}{2}}Q_{F_{i}}^{(N+m-2i)(l_{1}+\cdots l_{i})}
\prod_{i=\frac{N+m-1}{2}+1}^{N-1}Q_{F_{i}}^{(2i-N-m)(l_{i+1}+\cdots l_{N})}\,,\,\,N+m=\mbox{odd}\,\\ \nn
Q_{B}^{\sum_{i=1}^{N}l_{R_{i}}}
\,\prod_{i=1}^{\frac{N+m}{2}-1}Q_{F_{i}}^{(N+m-2i)(l_{1}+\cdots l_{i})}
\prod_{i=\frac{N+m}{2}+1}^{N-1}Q_{F_{i}}^{(2i-N-m)(l_{i+1}+\cdots l_{N})}\,,\,\,N+m=\mbox{even}\,.
\end{cases}
\label{product}
\eea

\section{Nekrasov's conjecture and the field theory limit} \label{conjecture}
Ever since the work of Seiberg and Witten \cite{SW}, it has been a challenge to reproduce their results using instanton calculus. In \cite{DKM,Dorey:1996bf,Flume:2001kb,Hollowood:2002ds}, this calculus is used to express the coefficients ${\cal F}_k$ of the $k$-instanton contributions to the prepotential as integrals over the moduli space of instantons which can be evaluated using localization techniques. These calculations become feasible when the integral localizes to a finite number of points. In \cite{Nekrasov} this was achieved by a certain deformation of the integrand by parameters $\epsilon_1$ and $\epsilon_2$. While the $\epsilon_{1,2} \rightarrow 0$ limit cannot be taken in the individual integrals $Z_k(\epsilon_1,\epsilon_2)$, \cite{Nekrasov} assembles 
these into an infinite sum 
\begin{eqnarray}
{\cal Z}(\varphi,\epsilon_1,\epsilon_2)=1+\sum_{k=1}^{\infty}Z_k(\epsilon_1,\epsilon_2)\varphi^k \,,
\end{eqnarray}
from which a generating function ${\cal F}(\varphi,\epsilon_1,\epsilon_2)=\sum_{k=1}^\infty {\cal F}_k(\epsilon_1,\epsilon_2)\varphi^k$ can be extracted via
\begin{eqnarray} \label{generating}
{\cal Z}=\exp(-\frac{1}{\epsilon_1 \epsilon_2}{\cal F}) \,,
\end{eqnarray}
such that $\lim_{\epsilon_{1,2} \rightarrow 0}{\cal F}_k(\epsilon_1,\epsilon_2)={\cal F}_k$. Somewhat surprisingly, ${\cal Z}$ has physical significance at finite $\epsilon_1=-\epsilon_2=\hbar$
\footnote{$\hbar$, following Nekrasov's notation, denotes an arbitrary constant.} as well. For this choice of parameters, \cite{Nekrasov} derives the following result (in our notation)
\begin{eqnarray}
{\cal Z}=\sum_{R_{1,\cdots,N}}\varphi^{l_{R_{1}}+\cdots + l_{R_{N}}}
\prod_{l,n=1}^{N}\prod_{i,j=1}^{\infty}\frac{a_{ln}+\hbar(\mu_{l,i}-\mu_{n,i}+j-i)}{a_{ln}+\hbar(j-i)}\,.
\end{eqnarray}
The sum over $R_{1,\cdots,N}$ runs over Young tableaux, as there is a $1:1$ correspondence between an ordered $N$-tupel of Young tableaux and the points at which the deformed integrals localize.
As conjectured in \cite{Nekrasov} and shown in \cite{firstpaper} in the case of $SU(2)$, this expression reproduces the field theory limit of the topological string partition function, for a particular choice of fibration of the resolved $A_n$ geometry over $\IP^1$, with the parameter $\hbar$ acquiring the role of the string coupling. 

In \cite{Nekrasov}, it was further conjectured that the following simple modification of this expression in fact reproduces the {\it complete} string partition function,
\bea
Z_{Nekrasov}:=\sum_{R_{1,\cdots,N}}\varphi^{l_{R_{1}}+\cdots +l_{R_{N}}}
\prod_{l,n=1}^{N}\prod_{i,j=1}^{\infty}\frac{\mbox{sinh}\frac{\beta}{2}(a_{ln}+\hbar(\mu_{l,i}-\mu_{n,i}+j-i))}{\mbox{sinh}\frac{\beta}{2}(a_{ln}+\hbar(j-i))}\,.
\eea
This conjecture was also verified, again in the $SU(2)$ case, in \cite{firstpaper}.

In this section, we wish to extend the verification of Nekrasov's conjecture to the general $SU(N)$ case. The calculation goes through almost exactly as in the $SU(2)$ case.

In \cite{firstpaper}, we noted that using the definition of ${\cal W}_{R}(q)$ 
and the following identity
\bea
\prod_{1\leq i<j<\infty}\frac{[\mu_{i}-\mu_{j}+j-i]}{[j-i]}=\prod_{1\leq i<j\leq d(\mu)}\frac{[\mu_{i}-\mu_{j}+j-i]}{[j-i]}\,\prod_{i=1}^{d(\mu)}\prod_{v=1}^{\mu_{i}}\frac{1}{[v-i+d(\mu)]}\,,
\eea
we have, with $q=e^{-\beta\hbar}$,
\bea
{\cal W}_{R}^{2}(q)=2^{-2l_{R}}q^{\kappa_{R}/2}\prod_{i,j=1}^{\infty}
\frac{\mbox{sinh}\frac{\beta\hbar}{2}(\mu_{i}-\mu_{j}+j-i)}{\mbox{sinh}\frac{\beta\hbar}{2}(j-i)}  \,.
\eea
Furthermore,
\bea
\prod_k (1-q^{k}Q)^{-2C_{k}(R_{r},R_{s}^{T})}&=&Q^{-l_{R_{r}}-l_{R_{s}}}\,2^{-2(l_{R_{r}}+l_{R_{s}})}\,q^{-\frac{1}{2}(\kappa_{R_{r}}-\kappa_{R_{s}})}\\ \nn
&&\prod_{l\neq n,i,j}\frac{\mbox{sinh}\frac{\beta}{2}(a_{ln}+\hbar(\mu_{l,i}-\mu_{n,j}+j-i))}{\mbox{sinh}\frac{\beta}{2}(a_{ln}+\hbar(j-i))}\,,
\eea
where $l,n \in \{r,s\}$, $i,j\geq 1$, $Q=e^{-\beta a_{r,s}}$.
The above two identities imply, using $\sum_{i<j}(\kappa_{i}-\kappa_{j})=\sum_{i=1}^{N}(N-2i+1)\kappa_{i}$,
\bea
\widehat{Z}^{(m)}&=&\sum_{R_{1,\cdots,N}}\prod_{i=1}^{N}Q_{b_{i}}^{l_{R_{i}}}\,
M^{(m)}(q,R_{i})2^{-2N(l_{1}+\cdots+l_{N})}\,
\prod_{i=1}^{N-1}Q_{i}^{-(N-i)(l_{1}+\cdots+l_{i})-i(l_{i+1}+\cdots+l_{N})}\,\\ \nn
&&q^{-\frac{1}{2}\sum_{i=1}^{N}(N-2i)\kappa_{i}}\prod_{l,n,i,j}\frac{\mbox{sinh}\frac{\beta}{2}(a_{ln}+\hbar(\mu_{l,i}-\mu_{n,j}+j-i))}{\mbox{sinh}\frac{\beta}{2}(a_{ln}+\hbar(j-i))}\,.
\eea
Using (\ref{product}) and (\ref{sunframing}) we see that for 
$m=0$,
\bea
\widehat{Z}^{(0)}=\sum_{R_{1,\cdots,N}}\varphi^{l_{1}+\cdots+l_{N}}\,\prod_{l,n,i,j}\frac{\mbox{sinh}\frac{\beta}{2}(a_{ln}+\hbar(\mu_{l,i}-\mu_{n,j}+j-i))}{\mbox{sinh}\frac{\beta}{2}(a_{ln}+\hbar(j-i))}\,,
\eea
where
\bea \nn
\varphi&=&\frac{Q_{B}}{2^{2N}D(Q_{F_{i}})}\,,\\ \nn
\eea
\begin{eqnarray}
D(Q_{F_{i}})=
\begin{cases}
\prod_{i=1}^{\frac{N-1}{2}}Q_{F_{i}}^{i}\prod_{i=\frac{N-1}{2}+1}^{N-1}Q_{F_{i}}^{N-i}\,\,\,,N=\mbox{odd} \nn \\
\prod_{i=1}^{\frac{N}{2}-1}Q_{F_{i}}^{i}\prod_{i=\frac{N}{2}+1}^{N-1}Q_{F_{i}}^{N-i}\,\,\,,N=\mbox{even}.\nn
\end{cases}
\eea
Taking the field theory limit
\bea
Q_{B}=(-1)^{N-m}(\frac{\beta\Lambda}{2})^{2N}\,,\,Q_{F_{j}}=e^{-\beta a_{j,j+1}}\,,\,\beta\rightarrow 0\,,
\eea
we get
\bea
{\cal Z}^{(m)}=\sum_{R_{1,\cdots,N}}(\frac{\Lambda}{2})^{2N(l_{1}+\cdots+l_{N})}\,\prod_{l,n,i,j}\frac{a_{ln}+\hbar(\mu_{l,i}-\mu_{n,j}+j-i)}{a_{ln}+\hbar(j-i)} \,.
\eea
Evaluating the sum above upto representations of combined length $k$ yields, by invoking (\ref{generating}), the instanton coefficients ${\cal F}_k$ of ${\cal N}=2$ $SU(N)$ gauge theory. This evaluation is performed in equations (3.23)
 and (3.24) of \cite{Nekrasov}, and the results coincide with previous work employing Seiberg-Witten techniques \cite{D'Hoker:1996nv}.

\section*{Acknowledgments}
AI would like to thank Marcos Mari\~no and Cumrun Vafa for valuable 
discussions. The research of AI was supported by NSF award NSF-DMS/00-74329. The 
research of AK was supported by the Department of Energy under contract number DE-AC03-76SF00515.

\section*{Appendix}

\subsection*{$SU(N)$ geometries}
In this section, we would like to sketch the origins of the diagrams encoding the $A_{n-1}$ fibrations over $\IP^1$ that we study in this paper. While not self contained, we hope that we will give the reader with a passing familiarity with toric geometry a clearer understanding of how these diagrams arise.

$A_{n-1}$ singularities are of the form $\IC^2/\IZ_n$. The corresponding toric diagram is depicted in \figref{an}.

\begin{figure}[h]
\begin{center}\leavevmode\epsfbox{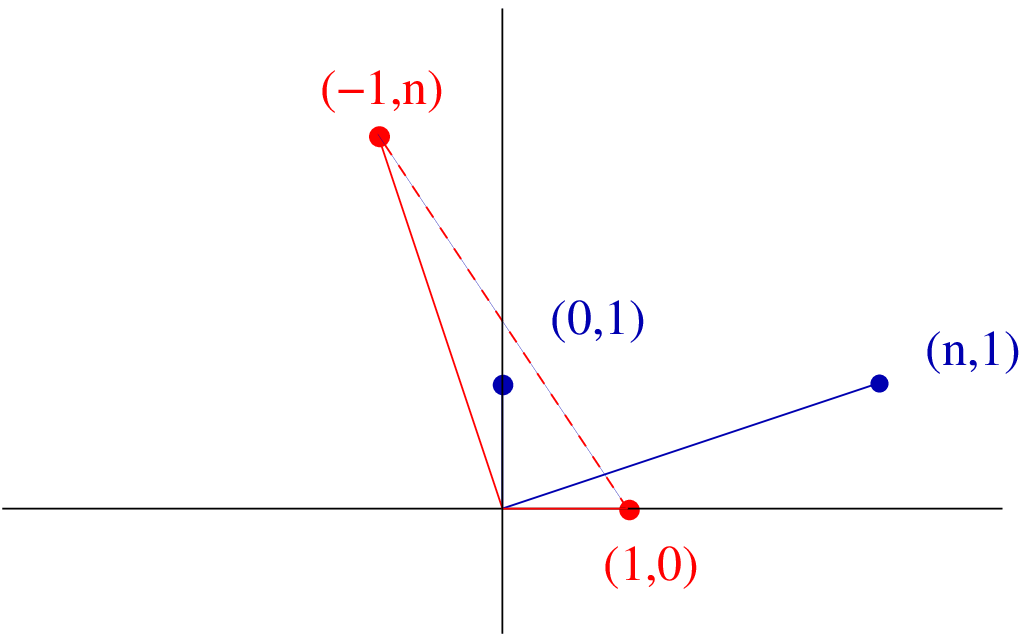}\end{center} 
\caption{\small The fan for $A_{n-1}$ (in blue), and its dual fan (in red). \label{an}}
\end{figure}

We can read off the coordinate ring of the toric variety from the dual fan. It is given by
\begin{eqnarray}
\IC[x,\frac{y^n}{x},y]&=& \IC[a,b,c]/(ab-c^n) \;,
\end{eqnarray}
which we recognize as the coordinate ring of $\IC^2/\IZ_n$. The fact that the corresponding variety is singular is encoded in the toric diagram in the fact that the single 2-cone comprising the fan is not generated by (part of) a basis of the lattice: rather than being generated by maximally two vectors, the fan is generated by the $n+1$ vectors $\{(1,0), (1,1), \ldots,(n,1) \}$. Subdividing the fan as depicted in \figref{anresolved} yields the toric diagram for the resolution of the $A_{n-1}$ singularity.

\begin{figure}[h]
\begin{center}\leavevmode\epsfbox{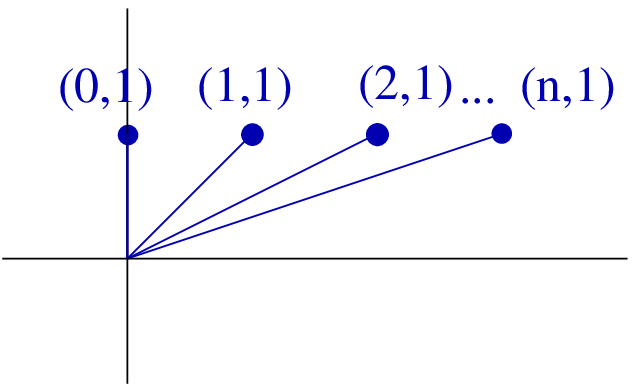}\end{center} 
\caption{\small The resolved $A_{n-1}$ geometry. \label{anresolved}}
\end{figure}

We now want to fiber these geometries over $\IP^1$, the toric diagram of which is depicted in \figref{p1}.

\begin{figure}[h]
\begin{center}\leavevmode\epsfbox{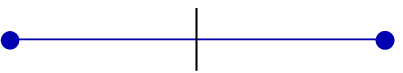}\end{center} 
\caption{\small The fan for $\IP^1$. \label{p1}}
\end{figure}

To this end, we embed \figref{anresolved} in a 3 dimensional lattice. To 
preserve the Calabi-Yau condition, we only add cones which are generated by 
vectors ending on the plane through the point $(0,1,0)$ and parallel to the 
$xz$ plane. In the diagrams in \figref{fibrations}, we omit the y direction. 
Adding any two cones such that their projection onto the $z$-axis yields the 
toric diagram of $\IP^1$ yields the desired geometry. The specific choice of 
cones determines how the resolved $A_{n-1}$ singularity is fibered over the 
$\IP^1$.

In \figref{fibrations}, we present the fans for fibrations of resolved $A_2$ over $\IP^1$, and the corresponding web diagrams. Note that this comprises all possible choices. If we move the vector $(a,1)$ further to the right than the $(2,1)$ position that yields the geometry with divisor $\IF_2$ and $\IF_4$, we obtain a space with more than two compact divisors. This is evident e.g. from the fact that the external legs of the web diagram start crossing past this point. On the other hand, if we move the vector further to the left than the (-1,0) position of the $\IF_1-\IF_1$ geometry, we reproduce fibrations already considered.

\begin{figure}[h]
\begin{center}
\leavevmode\epsfbox{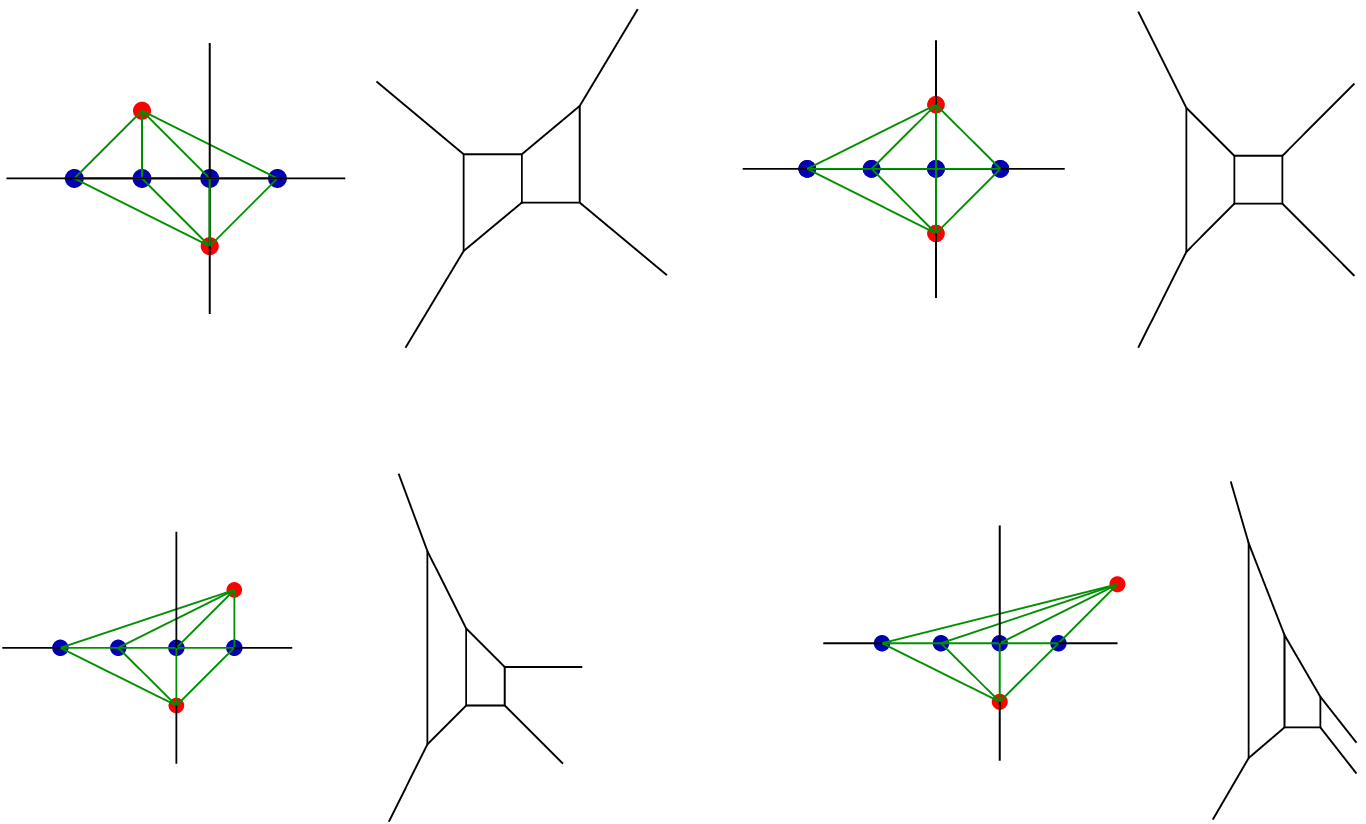}
\end{center} 
\caption{\small $A_2$ fibrations over $\IP^1$.}\label{fibrations}
\end{figure}

The considerations for the general case are completely analogous. Note that the web diagrams can also be obtained by gluing together the web diagrams of local Hirzebruch surfaces $\IF_k$, where the $k$'s of adjacent surfaces differ by $2$. We label the fibrations by an integer $m$, where $m=N$ denotes the geometry with the sequence $\{\IF_2,\IF_4,\ldots\}$ of divisors, and count downwards. For the case $N=2$, this reproduces the conventional labelling for Hirzebruch surfaces.

\subsection*{$SU(N)$ geometries and the $5d$ Chern-Simons coefficient}

The label $m$ we use to distinguish the various fibrations for a given $N$ is related to the triple intersection number of divisors and as such has a physical significance in the $5d$ theory one obtains by considering M-theory compactification,
instead of type IIA, on the CY3-folds we have been considering \cite{5dpapers, morrisonseiberg}. The 
5-dimensional theory has a prepotential with a cubic term. This
cubic term arises from the Chern-Simons term $\mbox{Tr}(A\wedge F\wedge F)$,
where $A$ is the gauge field and $F$ its field strength,
in the corresponding Lagrangian. The coefficient of this term, with appropriate normalization, is an integer called the Chern-Simons coefficient. From the 
CY point of view the cubic term
in the prepotential arises from the triple intersection numbers as follows \cite{morrisonseiberg,Iqbal:2002ep}. Let $S_{i}(m)$ be the various divisors which in our 
case are either even or odd Hirzebruch surfaces depending on $N+m$ even or odd,

\bea
S_{i}(m) \in \{\IF_{m+2-N},\IF_{m+4-N},\ldots,
\ff_{m+2r-N},\ldots,\ff_{m+N-2}\}\,.
\eea
Define $S(m)=\sum_{i=1}^{N-1}(\phi_{i+1}+\cdots \phi_{N})S_{i}$ where 
$\phi_{i}$, in the $5d$ theory,
parametrize the Coulomb branch moduli space. Then
\bea
S^{3}=\sum_{i,j,k}(\phi_{i+1}+\cdots\phi_{N})(\phi_{j+1}+\cdots +\phi_{N})(
\phi_{k}+\cdots +\phi_{N})\,(S_{i}\cdot S_{j}\cdot S_{k})
\eea
is such that
\bea
S^{3}(m)=\frac{1}{2}\sum_{i,j}|\phi_{i}-\phi_{j}|^3+m\sum_{i}\phi_{i}^3\,.
\eea
Thus the term we are using to label the geometries is exactly the
Chern-Simons coefficient of the $5d$ theory.

\end{document}